\documentclass[twocolumn]{aastex} 

\usepackage{style_paper}

\received{November 21, 2024}
\accepted{February 3, 2025}

\begin{document}
\title{exoALMA. VI. Rotating under Pressure: \\ \vspace{0.1cm}
     \text{Rotation curves, azimuthal velocity substructures, and pressure variations}}

\author[0000-0002-0491-143X]{Jochen Stadler}
\affiliation{Universit\'{e} C\^{o}te d'Azur, Observatoire de la C\^{o}te d'Azur, CNRS, Laboratoire Lagrange, France}

\correspondingauthor{Jochen Stadler}
\email{jochen.stadler@oca.eu}

\author[0000-0002-7695-7605]{Myriam Benisty}
\affiliation{Universit\'{e} C\^{o}te d'Azur, Observatoire de la C\^{o}te d'Azur, CNRS, Laboratoire Lagrange, France}
\affiliation{Max-Planck Institute for Astronomy (MPIA), Königstuhl 17, 69117 Heidelberg, Germany}

\author[0000-0002-7501-9801]{Andrew J. Winter}
\affiliation{Universit\'{e} C\^{o}te d'Azur, Observatoire de la C\^{o}te d'Azur, CNRS, Laboratoire Lagrange, France}
\affiliation{Max-Planck Institute for Astronomy (MPIA), Königstuhl 17, 69117 Heidelberg, Germany}

\author[0000-0001-8446-3026]{Andr\'es F. Izquierdo} 
\affiliation{NASA Hubble Fellowship Program Sagan Fellow}
\affiliation{Department of Astronomy, University of Florida, Gainesville, FL 32611, USA}
\affiliation{Leiden Observatory, Leiden University, P.O. Box 9513, NL-2300 RA Leiden, The Netherlands}
\affiliation{European Southern Observatory, Karl-Schwarzschild-Str. 2, D-85748 Garching bei M\"unchen, Germany}

\author[0000-0003-4663-0318]{Cristiano Longarini} 
\affiliation{Institute of Astronomy, University of Cambridge, Madingley Road, CB3 0HA, Cambridge, UK}
\affiliation{Dipartimento di Fisica, Universit\`a degli Studi di Milano, Via Celoria 16, 20133 Milano, Italy}

\author[0000-0002-5503-5476]{Maria Galloway-Sprietsma}
\affiliation{Department of Astronomy, University of Florida, Gainesville, FL 32611, USA}

\author[0000-0003-2045-2154]{Pietro Curone} 
\affiliation{Dipartimento di Fisica, Universit\`a degli Studi di Milano, Via Celoria 16, 20133 Milano, Italy}
\affiliation{Departamento de Astronom\'ia, Universidad de Chile, Camino El Observatorio 1515, Las Condes, Santiago, Chile}

\author[0000-0003-2253-2270]{Sean M. Andrews}
\affiliation{Center for Astrophysics | Harvard \& Smithsonian, Cambridge, MA 02138, USA}

\author[0000-0001-7258-770X]{Jaehan Bae}
\affiliation{Department of Astronomy, University of Florida, Gainesville, FL 32611, USA}

\author[0000-0003-4689-2684]{Stefano Facchini}
\affiliation{Dipartimento di Fisica, Universit\`a degli Studi di Milano, Via Celoria 16, 20133 Milano, Italy}

\author[0000-0003-4853-5736]{Giovanni Rosotti} 
\affiliation{Dipartimento di Fisica, Universit\`a degli Studi di Milano, Via Celoria 16, 20133 Milano, Italy}

\author[0000-0003-1534-5186]{Richard Teague}
\affiliation{Department of Earth, Atmospheric, and Planetary Sciences, Massachusetts Institute of Technology, Cambridge, MA 02139, USA}



\author[0000-0001-6378-7873]{Marcelo Barraza-Alfaro}
\affiliation{Department of Earth, Atmospheric, and Planetary Sciences, Massachusetts Institute of Technology, Cambridge, MA 02139, USA}


\author[0000-0002-2700-9676]{Gianni Cataldi} 
\affiliation{National Astronomical Observatory of Japan, 2-21-1 Osawa, Mitaka, Tokyo 181-8588, Japan}

\author[0000-0003-3713-8073]{Nicolás Cuello} 
\affiliation{Univ. Grenoble Alpes, CNRS, IPAG, 38000 Grenoble, France}

\author[0000-0002-1483-8811]{Ian Czekala} 
\affiliation{School of Physics \& Astronomy, University of St. Andrews, North Haugh, St. Andrews KY16 9SS, UK}



\author[0000-0003-4679-4072]{Daniele Fasano} 
\affiliation{Universit\'{e} C\^{o}te d'Azur, Observatoire de la C\^{o}te d'Azur, CNRS, Laboratoire Lagrange, France}

\author[0000-0002-9298-3029]{Mario Flock} 
\affiliation{Max-Planck Institute for Astronomy (MPIA), Königstuhl 17, 69117 Heidelberg, Germany}

\author[0000-0003-1117-9213]{Misato Fukagawa} 
\affiliation{National Astronomical Observatory of Japan, 2-21-1 Osawa, Mitaka, Tokyo 181-8588, Japan}


\author[0000-0002-5910-4598]{Himanshi Garg}
\affiliation{School of Physics and Astronomy, Monash University, Clayton VIC 3800, Australia}

\author[0000-0002-8138-0425]{Cassandra Hall} 
\affiliation{Department of Physics and Astronomy, The University of Georgia, Athens, GA 30602, USA}
\affiliation{Center for Simulational Physics, The University of Georgia, Athens, GA 30602, USA}
\affiliation{Institute for Artificial Intelligence, The University of Georgia, Athens, GA, 30602, USA}

\author[0000-0003-1502-4315]{Iain Hammond} 
\affiliation{School of Physics and Astronomy, Monash University, VIC 3800, Australia}

\author[0000-0001-7641-5235]{Thomas Hilder} 
\affiliation{School of Physics and Astronomy, Monash University, VIC 3800, Australia}

\author[0000-0001-6947-6072]{Jane Huang} 
\affiliation{Department of Astronomy, Columbia University, 538 W. 120th Street, Pupin Hall, New York, NY, USA}

\author[0000-0003-1008-1142]{John~D.~Ilee} 
\affiliation{School of Physics and Astronomy, University of Leeds, Leeds, UK, LS2 9JT}


\author[0000-0001-7235-2417]{Kazuhiro Kanagawa} 
\affiliation{College of Science, Ibaraki University, 2-1-1 Bunkyo, Mito, Ibaraki 310-8512, Japan}

\author[0000-0002-8896-9435]{Geoffroy Lesur} 
\affiliation{Univ. Grenoble Alpes, CNRS, IPAG, 38000 Grenoble, France}


\author[0000-0002-2357-7692]{Giuseppe Lodato} 
\affiliation{Dipartimento di Fisica, Universit\`a degli Studi di Milano, Via Celoria 16, 20133 Milano, Italy}

\author[0000-0002-8932-1219]{Ryan A. Loomis}
\affiliation{National Radio Astronomy Observatory, 520 Edgemont Rd., Charlottesville, VA 22903, USA}


\author[0000-0002-1637-7393]{Francois Menard} 
\affiliation{Univ. Grenoble Alpes, CNRS, IPAG, 38000 Grenoble, France}

\author[0000-0003-4039-8933]{Ryuta Orihara} 
\affiliation{College of Science, Ibaraki University, 2-1-1 Bunkyo, Mito, Ibaraki 310-8512, Japan}

\author[0000-0001-5907-5179]{Christophe Pinte}
\affiliation{Univ. Grenoble Alpes, CNRS, IPAG, 38000 Grenoble, France}
\affiliation{School of Physics and Astronomy, Monash University, Clayton VIC 3800, Australia}

\author[0000-0002-4716-4235]{Daniel J. Price} 
\affiliation{School of Physics and Astronomy, Monash University, Clayton VIC 3800, Australia}




\author[0000-0003-1412-893X]{Hsi-Wei Yen} 
\affiliation{Academia Sinica Institute of Astronomy \& Astrophysics, 11F of Astronomy-Mathematics Building, AS/NTU, No.1, Sec. 4, Roosevelt Rd, Taipei 10617, Taiwan}

\author[0000-0002-3468-9577]{Gaylor Wafflard-Fernandez} 
\affiliation{Univ. Grenoble Alpes, CNRS, IPAG, 38000 Grenoble, France}

\author[0000-0003-1526-7587	]{David J. Wilner} 
\affiliation{Center for Astrophysics | Harvard \& Smithsonian, Cambridge, MA 02138, USA}

\author[0000-0002-7212-2416]{Lisa W\"olfer} 
\affiliation{Department of Earth, Atmospheric, and Planetary Sciences, Massachusetts Institute of Technology, Cambridge, MA 02139, USA}

\author[0000-0001-8002-8473	]{Tomohiro C. Yoshida} 
\affiliation{National Astronomical Observatory of Japan, 2-21-1 Osawa, Mitaka, Tokyo 181-8588, Japan}
\affiliation{Department of Astronomical Science, The Graduate University for Advanced Studies, SOKENDAI, 2-21-1 Osawa, Mitaka, Tokyo 181-8588, Japan}

\author[0000-0001-9319-1296	]{Brianna Zawadzki} 
\affiliation{Department of Astronomy, Van Vleck Observatory, Wesleyan University, 96 Foss Hill Drive, Middletown, CT 06459, USA}
\affiliation{Department of Astronomy \& Astrophysics, 525 Davey Laboratory, The Pennsylvania State University, University Park, PA 16802, USA}


\begin{abstract}
The bulk motion of the gas in protoplanetary disks around newborn stars is nearly Keplerian. By leveraging the high angular and spectral resolution of ALMA, we can detect small-scale velocity perturbations in molecular line observations caused by local gas pressure variations in the disk, possibly induced by embedded protoplanets.
This paper presents the azimuthally averaged rotational velocity and its deviations from Keplerian rotation (\dvphi{}) for the exoALMA sample, as measured in the \twCOfull{} and \thCOfull{} emission lines. The rotation signatures show evidence for vertically stratified disks, in which \thCO{} rotates faster than \twCO{} due to a distinct thermal gas pressure gradient at their emitting heights. We find \dvphi{}-substructures in the sample on both small ($\sim$10\,au) and large ($\sim$100\,au) radial scales, reaching deviations up to 15\% from background Keplerian velocity in the most extreme cases.  
More than 75\% of the rings and 80\% of the gaps in the dust continuum emission resolved in \dvphi{} are co-located with gas pressure maxima and minima, respectively. Additionally, gas pressure substructures are observed far beyond the dust continuum emission. For the first time, we determined the gas pressure derivative at the midplane from observations and found it to align well with the dust substructures within the given uncertainties. Based on our findings, we conclude that gas pressure variations are likely the dominant mechanism for ring and gap formation in the dust continuum.
\end{abstract}


\section{Introduction} \label{sec:intro}
The exceptional capabilities of the Atacama Large Millimeter/submillimeter Array (ALMA) have made it possible to measure the rotation of gas in protoplanetary disks and search for the kinematic footprints of embedded protoplanets \citep[see review by][]{Pinte_ea_PPVII}. The rotational velocity of the gas \vphi{} orbiting a star of mass $M_\star$ at a given cylindrical radius $R$ and height $z$ assuming centrifugal balance 
is given by
\begin{equation} \label{eq:vrot}
\begin{aligned}
\frac{v_\phi(R, z)^2}{R}=\frac{G M_\star R}{\left(R^2+z^2\right)^{3 / 2}}+\frac{1}{\rho_{\mathrm{gas}}} \frac{\partial P_{\mathrm{gas}}}{\partial R} +\frac{\partial \phi_{\mathrm{gas}}}{\partial R},
\end{aligned}
\end{equation}
where $\rho_{\mathrm{gas}}(R, z)$ is the gas density, $P_\mathrm{gas}(R, z)$ is the gas pressure, and $\phi_{\mathrm{gas}}(R, z)$ is the gravitational potential of the disk \citep[e.g.,][]{Takeuchi_Lin_2002}. The first term on the right-hand side of the equation describes the dominant contribution from the star. 

Observationally, one measures the rotation curve from molecular line emissions which occurs from a given altitude in the disk. Thus, the dependence of the rotation curve on $z$ can be inferred by measuring the emission layer height, as it has been demonstrated for several systems \citep[][]{Pinte_ea_2018a, Law_ea_2021b, Law_ea_2023, Paneque_ea_2023}. The third term becomes important if the disk is sufficiently massive \citep[$M_\mathrm{disk} \geq 0.1 M_\star$, e.g.,][]{Lodato_2007, Kratter_Lodato_2016}, in which case it will speed up the gas, in particular in the outermost regions of the disk enclosing the most of the disk's mass. In recent years, studies have leveraged this effect to determine the total disk mass \citep[][]{Veronesi_ea_2021, Veronesi_ea_2024, Lodato_ea_2023, Martire_ea_2024, Andrews_ea_2024}. 

The second term is the acceleration due to the gas pressure gradient. It is of particular importance as it is most sensitive to localized variations in the physical conditions of the gas. On a global scale, the pressure gradient decreases with radius, as do the gas density and temperature, which introduces sub-Keplerian rotation in the outermost disk radii. However, local vertical and radial modulations in temperature and density can result in both large and small-scale perturbations to the velocity curve and thus deviations from Keplerian rotation \citep[e.g.,][]{Rosenfeld_ea_2013, Rab_ea_2020,Andrews_ea_2024, Pezzotta_ea_2025}. Leveraging high spatial and spectral resolution ALMA observations, it is possible to measure these localized small-scale perturbations in the rotational velocity \citep{Teague_ea_2018b, Teague_ea_2018c, Rosotti_ea_2020}. Furthermore, the derivative of the midplane pressure determines the drift rate of dust particles \citep[][]{Whipple_1972, Weidenschilling_1977, Takeuchi_Lin_2002, Barriere_ea_2005, Birnstiel_ea_2010}. As the dynamics of dust particles depends on the relative coupling to the gas, gas pressure minima and maxima can create gaps and rings in the millimeter continuum emission, respectively \citep[e.g.,][]{Paardekooper_Mellema_2004, Ayliffe_ea_2012, Pinilla_ea_2012, Dipierro_ea_2015, Stadler_ea_2022}. The observed continuum substructures can therefore be linked to local deviations from Keplerian rotation \citep{Teague_ea_2018b, Rosotti_ea_2020}. 

Recently, \cite{Izquierdo_ea_2023} studied the kinematics of the MAPS sample \citep[]{Oeberg_ea_2021} and found that nine out of eleven continuum rings in the disks are co-located with pressure maxima, traced by deviations in \vphi{}. Earlier, \cite{Izquierdo_ea_2022} illustrated the correlation between pressure minima and line width minima when observed through optically thick molecular lines. This correlation allows for robust detections of gaps in the gas surface density when examined in conjunction with the azimuthal velocity structure.
These studies demonstrate that analyzing the rotation curves of disks can greatly enhance our understanding of the dynamics and evolution of planet-forming disks. Utilizing the exceptional data quality from the exoALMA\footnote{ \url{https://www.exoalma.com}} Large Program \citep[0.1\arcsec{}, 26 m/s;][]{Teague_exoALMA}, such analysis is especially valuable, as it provides insights into the (sub)structures, kinematics, and physical conditions of our sample of disks.

This paper presents the rotation curves for the 15 disks of the exoALMA sample, measured from \twCOfull{} and \thCOfull{} molecular line emission. Even though \CSfull{} has also been observed as part of the program, we focus our analysis on the CO lines since it is difficult to extract reliable \vphi{} for the CS line for most of the disks. We assess the evidence of vertical stratification and the deviations from Keplerian rotation in the sample. In particular, we aim to investigate whether substructures in the gas pressure cause the observed dust continuum substructures. The paper is structured as follows: Section~\ref{sec:data} illustrates the choice of molecular line data cubes studied and Section~\ref{sec:methods} explains the methods applied to the data. In Sections~\ref{sec:disk_prop}~\&~\ref{sec:vphi_var}, we present the results for the global disk properties and variations of the rotational velocity co-located with the dust substructures, respectively. Finally, we discuss our results in Section~\ref{sec:discussion} and draw our conclusions in Section~\ref{sec:concl}.
\section{Data} \label{sec:data}
The calibration and imaging of the \twCOfull{}, \thCOfull{}, and \CSfull{} molecular line data cubes used for the analysis in this work are presented in detail in \cite{Loomis_exoALMA}. For each molecule and disks there are three sets of continuum-subtracted image cubes. Namely, these are the Fiducial Images (beamsize=0.15\arcsec{}), High Resolution Images (beamsize$<0.15\arcsec{}$), and High Surface Brightness Sensitivity Images (beamsize=0.30\arcsec{}; see \cite{Teague_exoALMA} for details). This paper utilizes all three image sets tailored to the specific disk region and scientific objectives we intend to explore, which motivates the necessary angular resolution and sensitivity. Specifically, we must find a balance between achieving high sensitivity for the faint outer disks and high angular resolution for the structure of the inner disk. To address this issue, in Section\,\ref{sec:disk_prop}, we employ the High Surface Brightness Sensitivity Images to study the global disk properties on large scales, while in Section\,\ref{sec:vphi_var}, we use the High Resolution Images to investigate small scale deviations from Keplerian rotation in the innermost disk regions co-located with the dust continuum. To this end, we spatially divide our analysis into two disk regions (inner vs. outer), separated at 0.9$R_{\rm d, 90}$ where $R_{\rm d, 90}$ is the radius enclosing 90\,\% of the continuum flux \citep[see][]{Curone_exoALMA}. This separation essentially divides the disks into an inner region co-located with the continuum emission and an outer one without large (millimeter) dust grains and substructures.

In order to accurately resolve the small-scale variations of the rotational velocities in the inner regions of the disks co-located with the continuum substructures, we require an angular resolution of at least half the major axis of the gas beam (beam$_{\rm major}$) to encompass the radial width of a gap or ring in the continuum. This is achieved with 0.10\arcsec{}$<$\,beam$_{\rm major}$\,$\leq$0.15\arcsec{}, channel spacings of 100-200~m/s and a signal-to-noise ratio (SNR) of $\geq$5 in the CO moment maps. On the other hand, to robustly measure the rotational velocity across the entire radial extent of the disk, we choose larger beam sizes (0.15\arcsec{}-0.30\arcsec{}) and channel spacings (100-200~m/s) to achieve a high SNR to resolve the whole spatial extent of the disk. In Figure~\ref{fig:dvphi_var_resolutions}, we show that these different choices of angular resolutions do not negatively affect the analysis products over the scales of interest \citep[see also][]{Teague_exoALMA}. A table with the employed data cubes for each disk and analysis can be found in \cite{Loomis_exoALMA} and the dust continuum substructures and their properties are reported in \cite{Curone_exoALMA}.

\section{Methods} \label{sec:methods}
\subsection{Line centroid velocity extraction} \label{sec:moments}
To accurately determine the line-of-sight velocity $\upsilon_\mathrm{l.o.s.}$, a precise measurement of the centroid of the peak of the line profile at every pixel is required. We modeled each disk and molecular line using the code \verb|discminer| \citep{Izquierdo_ea_2021} which fits a Keplerian model to the local line profile for each channel of a molecular line data cube. The strength of \verb|discminer| lies in its ability to model both the upper (front) and lower (back) surface emission. This is important for mid- to high-inclined sources ($i\sim\!30\degree-60\degree$), where the line profile is generally double-peaked due to the contribution of both surfaces. With a model for both emission surfaces at hand, it is possible to disentangle them to get a more accurate fit to the centroid of the line peak of the front surface. This is of particular importance in regions of the disk where the emission of the back surface of the disk becomes brighter than the upper surface or overlaps with it. In those cases with a line profile showing two components, we fit a two-component bell function to the line profile, which is essentially a Gaussian with one additional parameter that adds flexibility in controlling the line slope. For disks that only show a single peaked line profile, we generally fit a single Gaussian to determine the line centroid. The procedure of this newly improved double-bell kernel is presented in detail in \cite{Izquierdo_exoALMA}, as well as an overview of the choice of applied moment maps for each disk and molecular line.

\subsection{Rotational velocity} \label{sec:vel_extract}
The line-of-sight velocity can be decomposed into its disk-frame cylindrical components \vphi{}, \vrad{} and \vz{}. The respective projection of these components along the line-of-sight is given by
\begin{equation}
    \upsilon_{\rm los}=\upsilon_{\phi, \text { proj }} + \upsilon_{r, \text { proj }} + \upsilon_{z, \text { proj }} + \upsilon_{\rm LSR} \label{eq:v_los}
\end{equation}
\begin{equation}
     \upsilon_{\phi, \text { proj }}=\upsilon_\phi \cos (\phi) \sin (i) \cdot sgn_\mathrm{rot} \label{eq:vphi_los}
\end{equation}
\begin{equation}
    \upsilon_{r, \text { proj }}=-\upsilon_r \sin (\phi) \sin (i) \label{eq:vrad_los}
\end{equation}
\begin{equation}
    \upsilon_{z, \text { proj }}=-\upsilon_z \cos (i) \label{eq:vz_los}
\end{equation}
where $\upsilon_{\rm LSR}$ is the systemic velocity, $\phi$ the polar angle of the disk measured anticlockwise from the red-shifted major axis, ${i\in[-90\degr, 90\degr]}$ the disk inclination, and $sgn_\mathrm{rot}$ denotes whether the disk rotates clockwise (+1) or anti-clockwise ($-1$) \citep[see][for the adopted geometric conventions]{Izquierdo_exoALMA}.

We follow the analytical approach in \cite{Izquierdo_ea_2023} to extract the individual components from the $\upsilon_\mathrm{los}$. First, we extract the geometric parameters, including the offset from the center of the disk, the inclination, the position angle, and the systemic velocity of each disk from the best-fit parameters obtained from the \twCO{} \verb|discminer| fits \citep{Izquierdo_exoALMA}. Having these parameters, we can project a 2D grid onto our line centroid velocity moment maps. Next, we infer the rotational velocity \vphi{} for a specified radial annulus by taking the azimuthal average of the absolute value of the centroid velocity map $\upsilon_{0}$ after subtracting $\upsilon_{{\rm LSR}}$. This assumes the rotational velocity to be azimuthally symmetric around the disk minor axis and dominant over the radial and vertical ones:
\begin{equation} \label{eq:vphi_dm}
    \upsilon_\phi = \frac{\psi}{4\sin{\frac{\psi}{4}}\sin{|i|}} \left<\left|\upsilon_{0} - \upsilon_{{\rm LSR}}\right| \right>_{\psi},
\end{equation}
where $\psi$ denotes the angular extent of the azimuthal section where the averages are computed \citep[see Appendix C in][for derivation]{Izquierdo_ea_2023}. We mask 30$\degr$ around both sides of the disk minor axis, due to uncertainties in the deprojection at locations where $\cos(\phi)$ tends to 0. We extract radial profiles of \vphi{} starting from one major beam size from the disk center in steps of 1/4 major beam size and report the uncertainties of the profiles as the standard deviation normalized by the square root of independent beams along each radial annulus \citep{Izquierdo_ea_2023}. Finally, we apply a Savitzky-Golay filter of first order with a window length of one major beam size for smoothing and visual clarity of the plots in the main text. The quantitative analysis, however, is conducted on the raw data.

The deviations from Keplerian circular motion \dvphi{} can then be obtained by subtracting the Keplerian rotation ${\delta\upsilon_{\phi}= \upsilon_{\phi}-\upsilon_{\rm k}}$ given by
\begin{equation} \label{eq:vkep}
    v_{\rm k} (R, z) = \sqrt{\frac{G M_\star R^2}{{(R^2 +z^2)^{3/2}}}}.
\end{equation}
Unless explicitly stated otherwise, we obtain the stellar mass from the best-fit parameter of the \verb|discminer| fits for each molecular line, which is available for all sources \citep[][]{Izquierdo_exoALMA}. However, it is important to note that these $M_\star$ values should not be viewed as a \textit{true} stellar mass since \verb|discminer| only prescribes a pure Keplerian disk model (see Eq.\,\ref{eq:vkep}) and does not specifically account for the pressure and self-gravity terms of Eq.\,\ref{eq:vrot}. This leads to differences on the order of $\sim\!5\%$ between these best-fit values and the true stellar masses, as shown for a subset of our sample in \cite{Longarini_exoALMA}. In the following, we define $M_\star$ in Equation~\ref{eq:vkep} as the \textit{kinematic} stellar mass and elaborate on this further in the discussion. 

The emission heights of CO molecules (that go into Eq.\,\ref{eq:vkep}) usually arise from $z/R\approx0.2-0.3$ in disks \citep{Law_ea_2023, Galloway_exoALMA}. The elevation $z$ of the upper and lower emission surface, above and below the disk midplane, was modeled as an exponentially tapered power-law:
\begin{equation} \label{eq:em_height}
z(R)=z_0\left(\frac{R}{100 \mathrm{au}}\right)^{p_z} \exp \left[-\left(\frac{R}{R_t}\right)^{q_t}\right],
\end{equation}
where $z_0$, $p_z$, $R_t$ and $q_t$ are fitting parameters. For the analysis in this paper, we use the best-fit parameters of the \verb|discminer| surface fits, compatible with our velocity analysis framework.

The extraction of the other two velocity components, \vrad{} and \vz{}, will be presented in a forthcoming paper.

\subsection{Pressure variations} \label{sec:press_var}
Both the stellar and self-gravity terms in Eq.\,\ref{eq:vrot} solely introduce a large-scale component to the rotational velocity, hence, if observed, small-scale substructures in \vphi{} can be attributed to pressure variations. 

The radial gradient of the pressure profile can be directly related to the observed deviation in Keplerian rotation \dvphi{}. For now, we neglect the disk self-gravity, which is subdominant to the pressure \citep{Andrews_ea_2024, Longarini_exoALMA}. However, it is important to note that we will consider it later when calculating the pressure gradient at the midplane in Sec. \ref{sec:disc_pmid}. Neglecting $\phi_{\mathrm{gas}}$ allows us to rearrange Eq.~\ref{eq:vrot} as
\begin{equation} \label{eq:dPdr}
\left.\frac{\partial \ln P(R, z)}{\partial \ln R}\right|_{z=z_\mathrm{emit}} = \frac{\upsilon_\phi^2 - \upsilon_{\rm k}^2}{c_\mathrm{s}^2} \approx 2\frac{\upsilon_{\rm k}^2}{c_\mathrm{s}^2}\,\frac{\delta \upsilon_\phi}{\upsilon_{\rm k}},
\end{equation}
where we have approximated \vphi{}$\,\simeq\,$\vkep{} to first order, and introduced the sound speed $c_\mathrm{s}$ as ${P=\rho_{\mathrm{gas}}c_\mathrm{s}^2}$. The sound speed is related to temperature by
\begin{equation}
    c_\mathrm{s}^2 (R, z) = \frac{k_\mathrm{B} T(R, z)}{\mu m_\mathrm{p}},
\end{equation}
where $k_\mathrm{B}$ the Boltzmann constant and $\mu = 2.3$ the mean molecular weight for the proton mass $m_\mathrm{p}$.

\begin{figure}[t]
    \centering
    \includegraphics[width=1\linewidth]{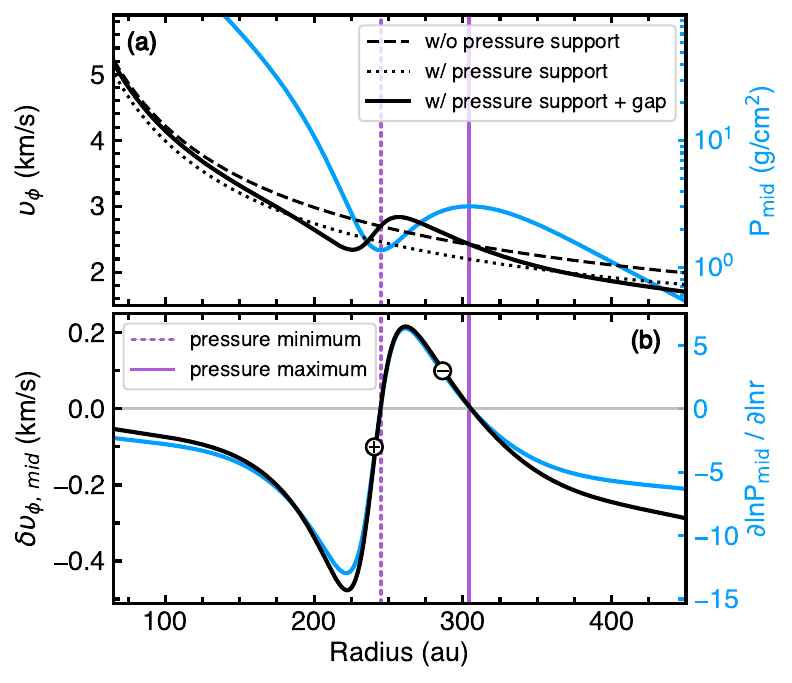}
    \caption{(a) Azimuthal velocities (left axis, black lines) and midplane pressure (right axis, blue lines), (b) along with the deviations from Keplerian rotation and the first radial derivative of the pressure, for a hydrodynamic model with an imposed Gaussian gap in the pressure at $R=240\,$au. The pressure minimum and maximum locations are marked with vertical purple dashed and full lines, respectively. The circles in (b) mark the sign of the radial derivative of \dvphi{} coinciding with the pressure minimum ($\oplus$) and maximum ($\ominus$), respectively.}
    \label{fig:vphi_press}
\end{figure}

To demonstrate how pressure gradient and azimuthal velocity (or its gradient) are related to each other, we set up a toy model. Specifically, we set the background disk density and temperature structures identical to the ones used for the HD~163296 disk as presented in \cite{Teague_ea_2019a}, and impose a Gaussian gap centered at $R=240$\,au, with a radial width of 50~au, and the maximum depth at the gap center of $90~\%$. Following \cite{Teague_ea_2019a}, the rotational velocity is numerically computed such that the disk is in centrifugal balance. Figure~\ref{fig:vphi_press} illustrates the expected variation of the rotational velocity for the disk having a Gaussian gap. We highlight that this toy model is for illustration purposes, demonstrating stronger pressure perturbations than those typically observed in our dataset.

In panel (a), we plotted the background rotation for a disk with (dotted black line) and without (dashed black line) pressure support, which corresponds to the first \& second term, or only the first term of Eq.\,\ref{eq:vrot}, respectively. From the model, we expect $\delta\upsilon_{\phi,\rm{mid}}$ and \dpmid{} at the midplane (subscript ``mid'') to be equal to zero at the location of pressure minima and maxima, as shown in Fig.~\ref{fig:vphi_press}\,b). This holds assuming that we know exactly the stellar mass. However, as pointed out in Sec.\,\ref{sec:vel_extract}, disentangling the various contributions to the rotational velocity (Eq.\,\ref{eq:vrot}) makes it difficult to constrain $M_\star$ precisely. This in turn results in a vertical shift in $\delta\upsilon_{\phi,\rm{mid}}$, thus we cannot accurately identify the pressure minima or maxima at locations of $\delta\upsilon_{\phi,\rm{mid}}$=0. Similar to \cite{Rosotti_ea_2020}, we circumvent this problem by investigating the sign of the radial derivative of \dvphi{}. Hence, around the location of a pressure minimum, we expect \dvphi{} to increase radially ($\oplus$), and the opposite for a pressure maximum ($\ominus$; compare to markers in Fig.\,\ref{fig:vphi_press}b). This diagnostic will be used to infer pressure minima and maxima presented in this work. In Section~\ref{sec:disc_pmid}, we will also infer the midplane pressure derivative for a subset of our sample.

\begin{figure*}[t]
    \centering
    \includegraphics[width=1\linewidth]{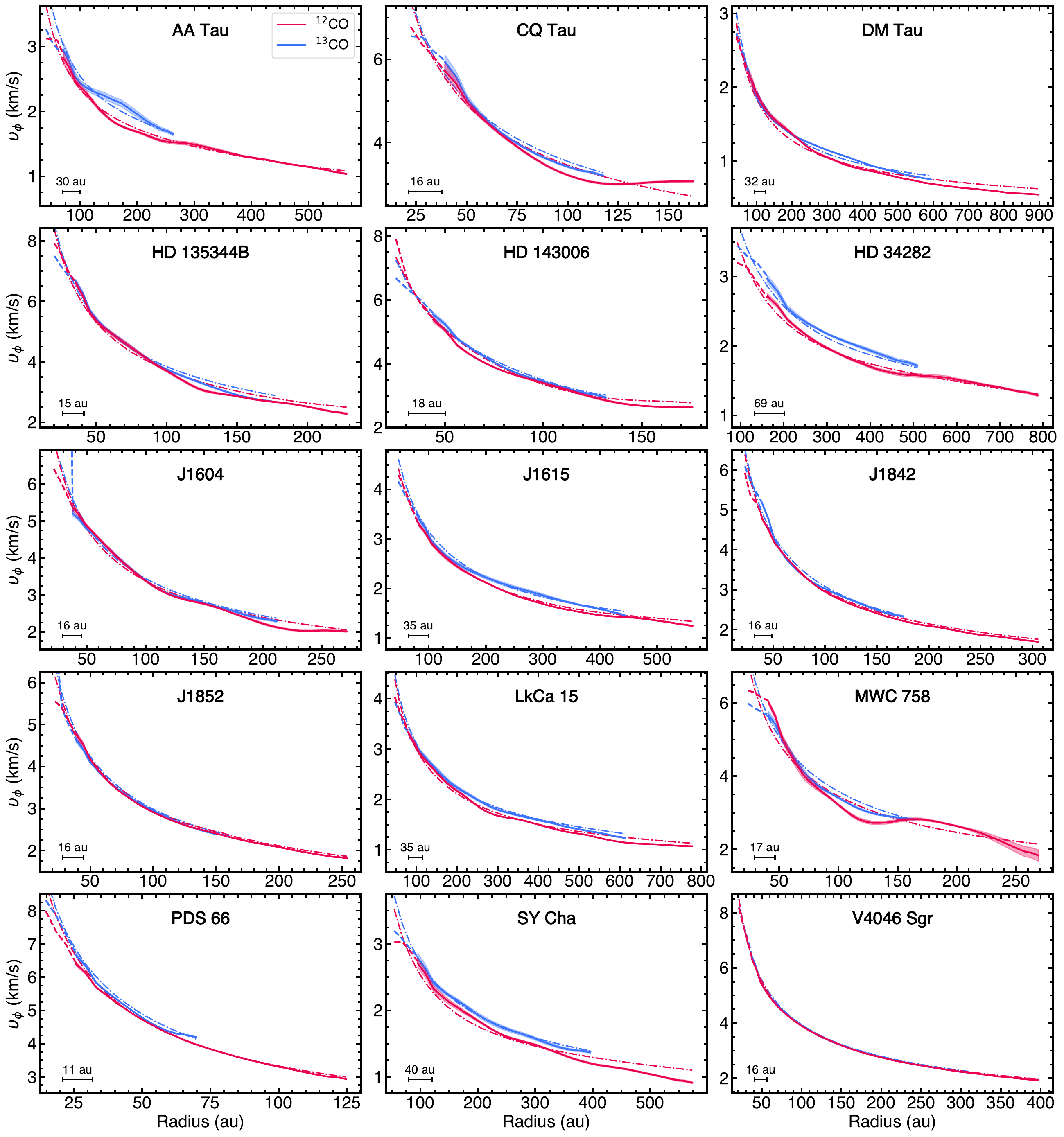}
    \caption{Rotation curves \vphi{}($R, z(R)$) measured from \twCO{} (red) and \thCO{} (blue), shown of all sources using the High Surface Brightness Sensitivity Images. For radii smaller than twice the beam size, the curves are plotted with dashed lines due to uncertainties in the velocity extraction. The colored shaded area of the lines shows the standard deviation within each extracted radial annulus usually on the order of $\approx10-50$ m/s. The fitted Keplerian rotation is plotted as a thin dashed-dotted line for each molecule respectively (Eq.~\ref{eq:vkep}). The beam size for both molecular lines is plotted in the lower left corner.}
    \label{fig:rot_curves}
\end{figure*} 

\subsection{Properties of \dvphi{} substructures} \label{sec:report_substr}
In Sec.\,\ref{sec:vphi_var}, we present \dvphi{} variations across the full disk extent, where numerous substructures due to pressure variations are present. However, we focus on the question of whether pressure substructures align with the continuum gaps and rings. Thus, we report quantitative properties of \dvphi{} substructures only when (1) they are co-located with continuum substructures. We further restrict our analysis to (2) axisymmetric continuum substructures (discarding crescents) at locations that are well resolved. The latter means, we only consider ``strong'' dust substructures that satisfy (3) a contrast $I_\mathrm{D}/I_\mathrm{B}<$0.8 between their gap ($I_{\rm D}$) and ($I_{\rm B}$) ring intensity, discarding very subtle gap-ring-pairs \citep[$I_\mathrm{D}/I_\mathrm{B}<$0.96 in][]{Curone_exoALMA}. Additionally, the continuum substructure must exhibit (4) a radially increasing (at dust gaps) or decreasing (at dust rings) \dvphi{}-profile for two consecutive data points. Finally, (5) the radial extent of the \dvphi{}-perturbation must exceed half of a major beam size. For \dvphi{} substructures satisfying all of the above five criteria, we measure their width as the radial extent between two data points ($\delta\upsilon_{\phi, \mathrm{min}}(r_{\rm min})$, $\delta\upsilon_{\phi, \mathrm{max}}(r_{\rm max})$) along which \dvphidr{} stays either positive or negative. The center between these two points then gives the radial location of the substructure, their amplitude is reported as $A=|$\dvphi{}$_\mathrm{, max}$-\dvphi{}$_\mathrm{, min}|$ and its uncertainty as $\Delta A=(\Delta\delta\upsilon_{\phi, \mathrm{max}}^2+\Delta\delta\upsilon_{\phi, \mathrm{min}}^2)^{1/2}$. We assume the uncertainties in the measurements of the radial location and width to be half the full width at half maximum (FWHM) of the major beam size. 

We comment on continuum substructures that do not satisfy one or more of the above criteria in Table~\ref{tab:non_substructures} of the Appendix. Our measurement of the \dvphi{} gap's amplitude, width, and radial location follows the approach presented in \cite{Yun_ea_2019}. We can thus use their derived relations to convert the gap \dvphi{}-amplitudes into estimates of a planet mass potentially driving the perturbation, as presented in Appendix \ref{app:planet_mass}.

\section{Global disk properties} \label{sec:disk_prop}
\subsection{Vertical thermal stratification} \label{subsec:vert_strat}
In Figure~\ref{fig:rot_curves}, we show the rotation curves for the molecular lines \twCOfull{} and \thCOfull{} for all 15 disks of the sample using the High Surface Brightness Sensitivity Images. We focus our analysis on these lines and omit the rotation curves of \CSfull{}, which has also been observed as part of the program since it is difficult to extract reliable \vphi{} for most of the disks due to a combination of limited angular resolution and SNR of the line. Nonetheless, the retrieved CS rotation curves can be found in Fig.~\ref{fig:rot_curves_CS} of the Appendix. 
As seen in Figure~\ref{fig:rot_curves}, we can infer \twCO{} and \thCO{} \vphi{} out to radial extents of several hundreds of au. For the largest disks in the sample, LkCa\,15 and DM\,Tau, we trace the \twCO{} rotation out to about 800\,au and 900\,au, respectively. 

\begin{figure*}
    \centering
    \includegraphics[width=1\linewidth]{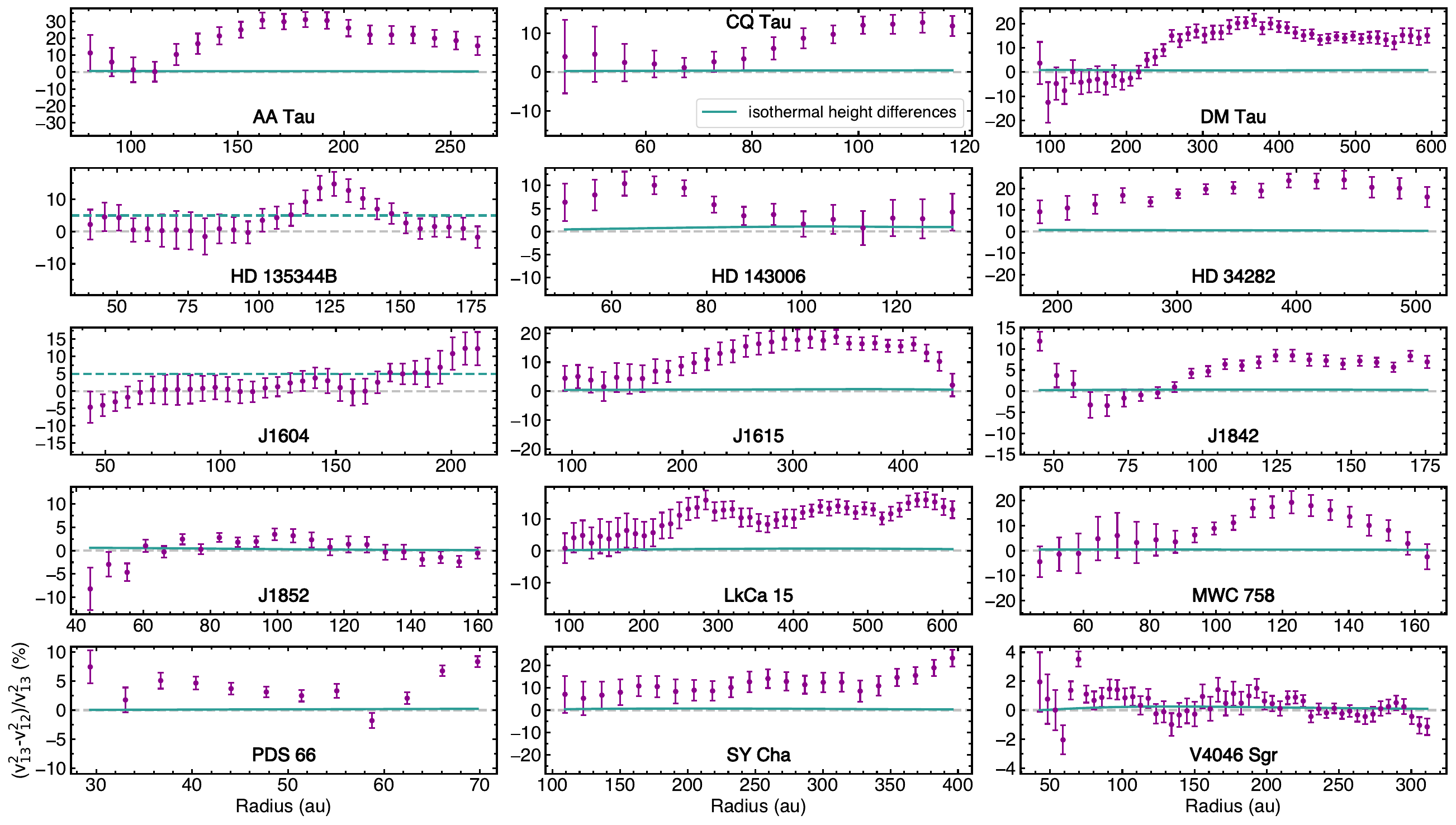}
    \caption{Level of vertical stratification between \twCO{} and \thCO{} rotation curves outside of two beam sizes from the center. The turquoise line displays the expected isothermal velocity shift solely due to the differences in emitting height (RHS Eq.\,\ref{eq:vert_strat}).}
    \label{fig:vert_strat}
\end{figure*} 

It is evident that the \twCO{} and \thCO{} rotation curves do not match. This discrepancy may arise from the fact that they trace distinct heights and stellar potentials in the disks, or from differences in their thermal pressure gradient at those heights (essentially stellar vs. pressure term in Eq.\,\ref{eq:vrot}). The earlier can be quantified following the approach of \cite{Martire_ea_2024}, who studied the effect of vertical \textit{thermal} stratification on rotation curves. First, we plot the relative differences of the squared rotational velocity for \twCO{} ($\upsilon_{\phi,12}$) and \thCO{} ($\upsilon_{\phi,13}$) in Fig.~\ref{fig:vert_strat}. Now, we hypothetically assume that the disks are vertically isothermal, thus the differences between both rotation curves should be solely due to their differences in emission height. This can be expressed as follows:
\begin{equation} \label{eq:vert_strat}
\frac{\upsilon_{\phi,13}^2-\upsilon_{\phi,12}^2}{\upsilon_{\mathrm{k}}^2}= -q_{\rm mid} \frac{\sqrt{1+z_{12}^2 / R^2}-\sqrt{1+z_{13}^2 / R^2}}{\sqrt{\left(1+z_{13}^2 / R^2\right)\left(1+z_{12}^2 / R^2\right)}},
\end{equation}
\citep[Eq.\,23,][]{Martire_ea_2024} where $q_{\rm mid}$ is the radial temperature power-law index at the midplane obtained from \cite{Galloway_exoALMA} and $z_{12}$ and $z_{13}$ denote the \verb|discminer| emission height for \twCO{} and \thCO{}, respectively. For the sources where no 2D temperature structure and thus no $q_{\rm mid}$ could be derived, we set $q_{\rm mid}=-0.32$ which is the mean of our sample. Lastly, for HD\,135344B and J1604 no emission surface could be extracted due to their face-on nature. For those cases, we estimate an upper limit of the velocity shift for the right-hand side of Eq.\,\ref{eq:vert_strat} following \cite{Martire_ea_2024} which assumes that for both CO molecules $z/R<$0.5 holds. Then the upper limit for the right-hand side becomes $<5\%$. This expected isothermal velocity shift due to differences in emitting height is plotted as a turquoise line in Figure~\ref{fig:vert_strat}. 

In Figure \ref{fig:vert_strat}, it is evident that for the majority of the sample, the gas at the emission surface traced by \thCO{} rotates faster than \twCO{}, and the differences are larger than can be explained by differences in gravity assuming a constant temperature. Therefore, we can infer that the thermal pressure gradient differs at distinct heights in the disks. In simpler terms, vertical thermal stratification is noticeable in most of the disks. It is particularly pronounced for AA\,Tau, DM\,Tau, HD\,34282, J1615, LkCa\,15, and SY\,Cha, the largest disks in the sample. The exceptions are J1852, V4046\,Sgr, and the face-on sources (HD\,135344B, HD\,143006, J1604), where the ratio is around or below the isothermal height differences line. These results suggest a dependency with orbital radius for this method, which we discuss later in Section~\ref{sec:disc_offset}.

\begin{figure*}[t]
    \centering
    \includegraphics[width=1\linewidth]{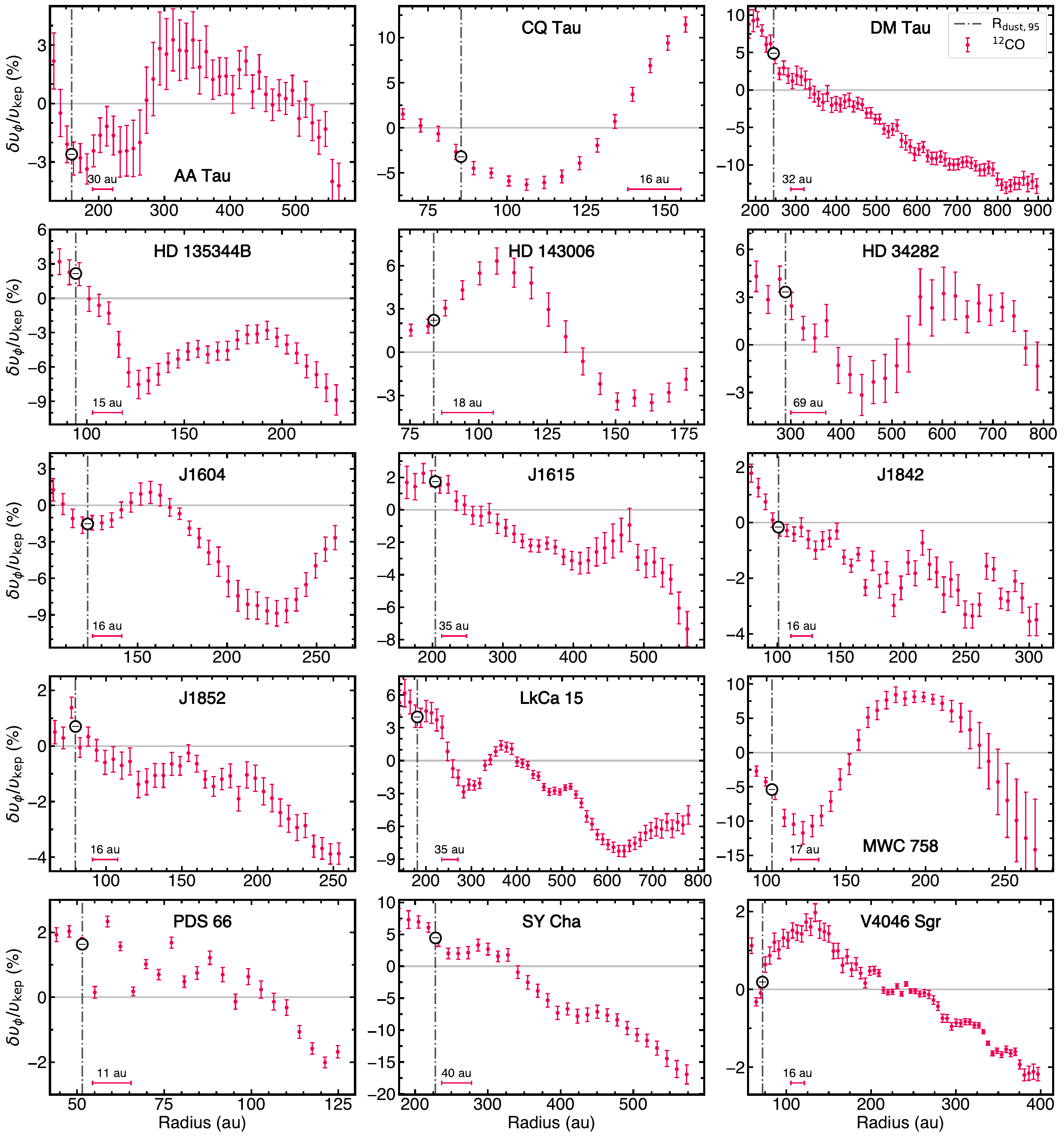}
    \caption{Radial profiles of \dvphi{} for \twCOfull{}{} of all sources focused on the region beyond the continuum substructures using the High Surface Brightness Sensitivity Images. The error bars show the standard deviation within each extracted radial annulus. The vertical dashed-dotted line indicates the radius that encompasses 95\% of the continuum emission, with the signs of the radial derivatives of \dvphi{} marked at this location. The beam size is shown in the lower left corner. The \thCO{} \dvphi{} radial profiles can be found in Fig.~\ref{fig:dvphi_large_13co} of the Appendix.}
    \label{fig:dvphi_large}
\end{figure*}

\subsection{Pressure drop-off in the outer disk} \label{subsec:dvphi_outer_disk}
Assuming the disks are in centrifugal balance, we expect \dvphi{} variations to be mostly driven by modulations in the underlying pressure structure. In this section, we focus on the \twCO{} molecular line to investigate the large-scale pressure modulations, as it offers higher sensitivity in the outermost regions of the disk. In Figure~\ref{fig:dvphi_large}, we present the \twCO{} radial \dvphi{} profiles showing the deviation from Keplerian rotation in the outer disk beyond the continuum, using the High Surface Brightness Sensitivity Images. The \thCO{} \dvphi{} profiles are presented in Fig.\,\ref{fig:dvphi_large_13co} of the Appendix, which show similar features as the \twCO{} profiles though not tracing as far out in radius. At first glance, one can see numerous substructures in the \dvphi{} profiles for nearly all of the disks. 

Most of the disks show an overall declining \dvphi{} slope and sub-Keplerian rotation in the outer disk regions. This is as expected due to the pressure fall off, since the temperature and the density of the disk decrease with radius. For a power-law density profile, this would introduce only a minor deviation of about 1-2\,\% \vkep{} \citep{Rosenfeld_ea_2013, Andrews_ea_2024}. However, if the density drops off exponentially \citep[e.g.][]{Lynden_Bell_Pringle_1974}, the disk would experience a more substantial slowdown of the gas rotation, which has been discussed in detail by \cite{Dullemond_ea_2020} (see their Sec.\,4). They showed that a sharp exponential cut-off in the disk density distribution ($\gamma>1$ for a Lynden-Bell \& Pringle $\Sigma$-profile), reminiscent of the disk's outer edge, leads to a strong negative pressure gradient and thus significant subkepler rotation. The most striking examples of this effect are DM\,Tau and SY\,Cha, where the \twCO{} Keplerian rotation slows down by more than 10\,\% and 15\,\% at their outermost radii, respectively, tracing their outer disk edges. Nevertheless, half of the sample only shows deviations on the order of a few percent Keplerian on large radial scales in their \twCO{} \dvphi{} profiles.

Our findings of sub-Keplerian rotation in the outer regions of the disk are consistent with the self-consistent modeling of the CO rotation curves presented in \cite{Longarini_exoALMA}. In their study, they fitted the exponential tapering radius, which indicates where the gas density and pressure fall off, and typically found this radius to be several hundred au closer in than the edge of the disk. The exceptions to the previously described behavior of \dvphi{} are CQ\,Tau and MWC\,758, who exhibit super-Keplerian rotation in their outer disk regions due to strong asymmetric and non-Keplerian motions, visible as prominent spiral morphology in their velocity residuals \citep{Izquierdo_exoALMA}. 

\section{Rotational velocity variations at dust substructures} \label{sec:vphi_var}
\subsection{\dvphi{} curves} \label{subsec:dvphi_inner_disk}
\begin{figure*}
    \centering
    \includegraphics[width=1\linewidth]{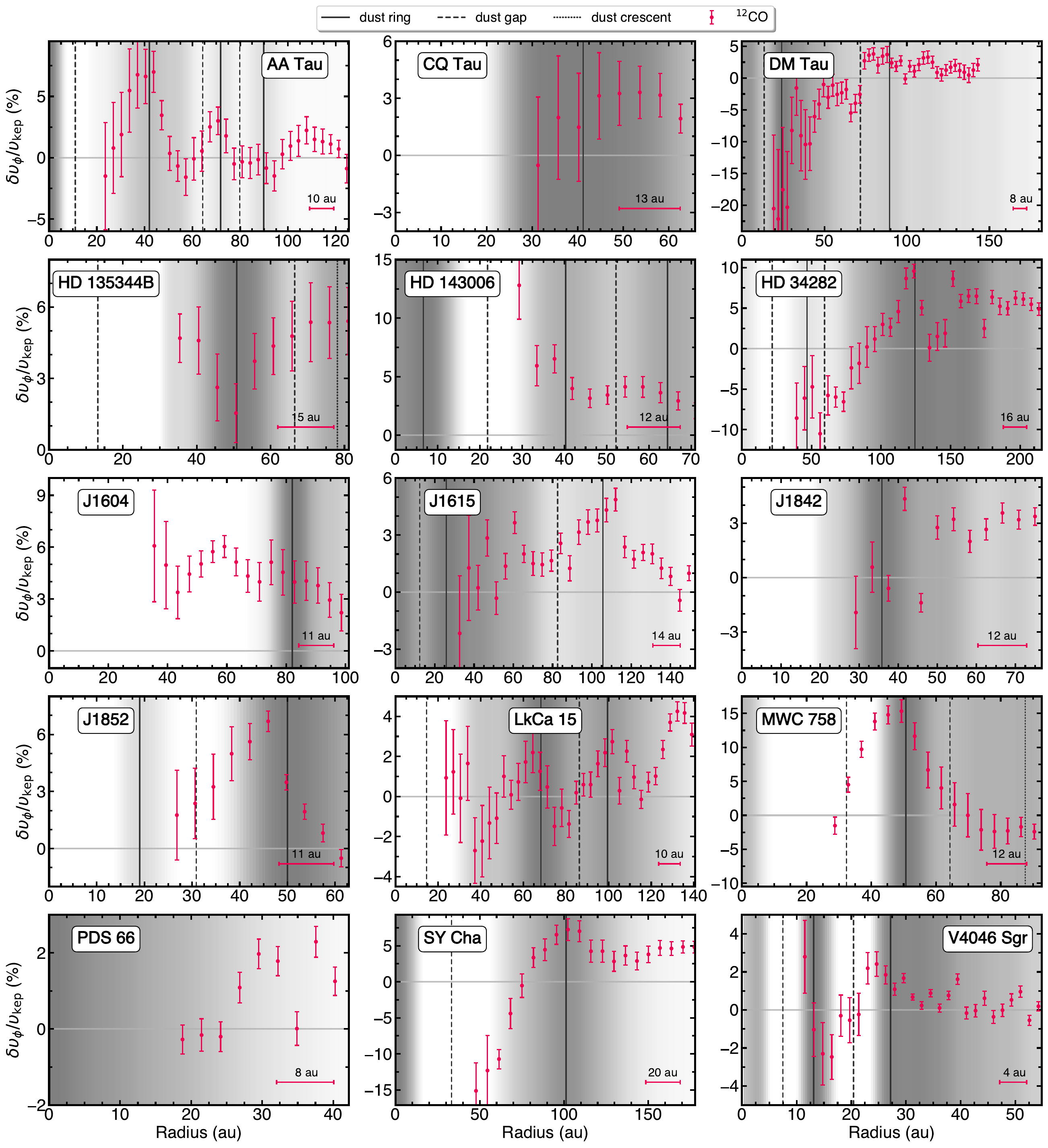}
    \caption{Radial profiles of \dvphi{} for \twCOfull{}{} of all sources focused on the region of the continuum substructures, using the High Resolution Images. The profiles are plotted starting at two beam sizes from the disk center and the error bars show the standard deviation of each bin. The gray background gradient highlights the \texttt{frank} radial profiles of the dust continuum emission normalized to its peak. The locations of continuum rings, gaps, and crescents are plotted in solid, dashed, and dotted vertical gray lines, respectively \citep[][]{Curone_exoALMA}. We expect dust rings and gaps to be co-located with a radially decreasing and increasing \dvphi{} profile, if caused by pressure maxima and minima, respectively. The beam size is plotted in the lower right corner. The \thCO{} \dvphi{} radial profiles can be found in Fig.~\ref{fig:dvphi_small_13co} of the Appendix.}
    \label{fig:dvphi_small}
\end{figure*}

In Figure~\ref{fig:dvphi_small}, we present the deviations from Keplerian rotation for the \twCO{} molecular line in the innermost disk regions coinciding with the dust continuum substructures \citep[][]{Curone_exoALMA}. The \thCO{} \dvphi{} profiles are shown of Fig.\,\ref{fig:dvphi_small_13co} of the Appendix. For both molecular lines, we use the High-Resolution Images (beam$_{\rm major}\leq 0.15\arcsec{}$ presented in \cite{Teague_exoALMA}) to access these radii and to radially resolve \vphi{} at the underlying dust substructures. Similarly to Fig.~\ref{fig:dvphi_large}, we also observe many \dvphi{}-substructures in the inner disk regions, which we report in Table\,\ref{tab:substructures} as outlined in section\,\ref{sec:report_substr}. We decided only to show the profiles starting at a radius of two major beam sizes from the disk center since we resolve the circumference of closer-in annuli only by less than 12 major beam sizes. This limited angular resolution leads to artificial sub-Keplerian motion and large uncertainties within each of those inner radial bins due to beam-smearing and velocity-mixing effects \citep[e.g.,][]{Andrews_ea_2024, Hilder_exoALMA} which are discussed in more detail in Sect.\,\ref{sec:discus_measure}. The outer plotting limit is set to $0.9\,R_{\rm dust, 90}$, which is the radius enclosing 90\,\% of the continuum emission.

It is apparent, that most of the transition disks (HD\,34282, LkCa\,15, MWC\,758, SY\,Cha) show radially increasing \dvphi{} profiles in their dust cavities, expected from density gaps at these locations. The exceptions of super-Keplerian velocities paired with radially decreasing \dvphi{} profiles for \twCO{} in the transition disks of J1604 and HD\,143006 can be attributed to highly non-Keplerian motions induced by warps and/or radial flows observed in these cavities \citep{Rosenfeld_ea_2014,Stadler_ea_2023}. For some disks, our angular resolution (number of independent beams along an annulus) and SNR are insufficient to assess their innermost cavities or dust substructures (e.g. CQ\,Tau, or V4046). 

\begin{figure}[t]
    \centering
    \includegraphics[width=1\linewidth]{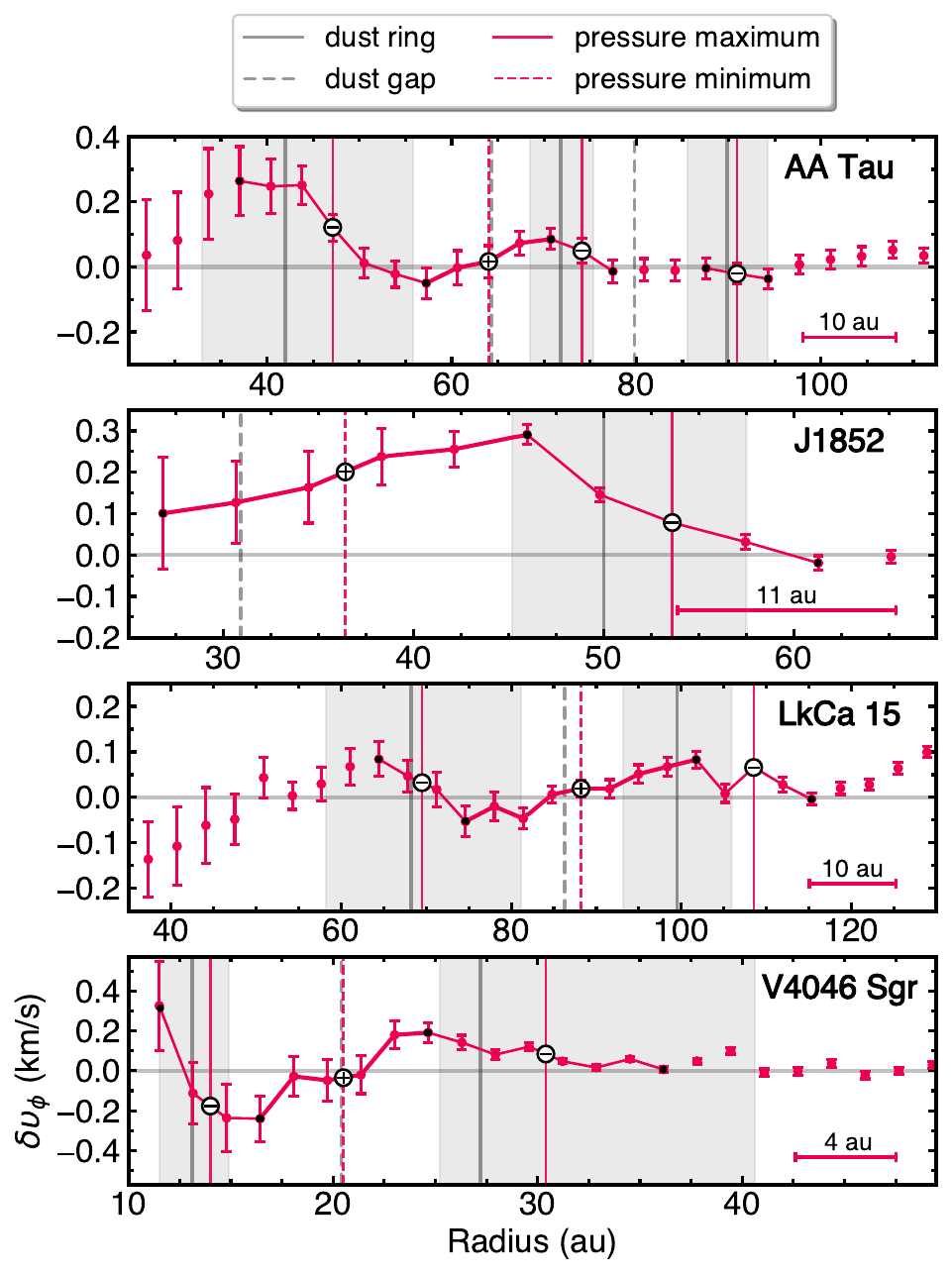}
    \caption{\twCO{} \dvphi{} profiles for selected sources focused on the region of the continuum emission with highlighted radial locations of pressure substructures (red vertical lines). From Fig.\,\ref{fig:vphi_press}, we expect dust rings and gaps to be co-located with negative ($\ominus$) and positive ($\oplus$) \dvphi{} gradients, if caused by pressure maxima and minima, respectively. The vertical grey shaded areas show the width of the continuum rings \citep[][]{Curone_exoALMA}. The other plot annotations are the same as in Fig.~\ref{fig:dvphi_small}.}
    \label{fig:dvphidr_small_selected}
\end{figure}

\subsection{Pressure variations} \label{sec:res_p_variations}
In this section, we investigate the pressure substructures of the disks by analyzing the radial derivative of \dvphi{} as discussed in Section\,\ref{sec:press_var}. Our goal is to determine whether the observed continuum gaps and rings align with pressure minima and maxima, respectively (refer to Fig.~\ref{fig:vphi_press}(c)). In Figure~\ref{fig:dvphidr_small_selected}, we present the \twCO{} \dvphi{}-profiles for four selected targets, where the sign of \dvphidr{} at the continuum substructures matches the theoretical expectations: \dvphidr{} has a negative sign at the location of the continuum ring and a positive sign at a gap, as highlighted via the signs on the solid curves in Fig.~\ref{fig:dvphidr_small_selected}. For dust substructures exhibiting this theoretical behavior, indicating a decrease in \dvphi{} within their continuum ring width (or an increase within gap widths), we conclude that the dust gaps and rings coincide with pressure minima and maxima, respectively. We then characterize these \dvphi{} substructures and report their properties in Table\,\ref{tab:substructures} following the procedure outlined in Section\,\ref{sec:report_substr}. We further comment on the \dvphi{}-substructures that, although co-located with continuum substructures, do not align with the theoretical expectation of pressure maxima/minima, or could not be assessed due to limited angular resolution in Table\,\ref{tab:non_substructures} of the Appendix. The latter substructures are not included in the total count of non-aligning \dvphi{}-substructures considered. We find that for \twCO{}, 16 out of 21 continuum rings and 10 out of 12 continuum gaps align with negative and positive \dvphidr{}, respectively. In the case of \thCO{}, \dvphidr{} is negative for 14 out of 17 rings and positive for 8 out of 10 gaps. Therefore, more than 75\% of dust continuum rings and 80\% of gaps in both \twCO{} and \thCO{} coincide with local maxima and minima in the gas pressure. We emphasize that these pressure substructures identified through \dvphi{} variations in gas kinematics are reflected in the line centroids, not in the line \textit{intensities}. Annotated \dvphi{}-profiles highlighting the reported substructures listed in Table\,\ref{tab:substructures} can be found in Figures\,\ref{fig:dvphi_substr_12co}~\&~\ref{fig:dvphi_substr_13co} of the Appendix.

Looking at large-scale \dvphi{} variations in Figure~\ref{fig:dvphi_large}, most of the $R_{\rm dust,95}$ coincide with negative radial \dvphi{}-gradients in \twCO{} (13 out of 15), which hints toward radial inward drift of dust particles, i.e. the dust extent is shrinking. On the other hand, for HD\,143006 and V4046\,Sgr the gradient is positive in both tracers, which could point to the co-location with a local pressure bump, i.e. a more stable dust outer edge. However, the \thCO{} velocities, expected to be better tracers of the dust distribution near the midplane, show fewer negative slopes (only 7 out of 15, see Fig.~\ref{fig:dvphi_large_13co}). This leads to ambiguity in some disks between \twCO{} and \thCO{}, where either slopes are decreasing or increasing, which will be further discussed in Section\,\ref{sec:disc_offset}. In the outermost disk, beyond the continuum, \dvphi{} is radially decreasing within the outermost beam, i.e. \dvphidr{} has a negative sign, for 10 out of 15 disks for \twCO{} and 8 out of 15 disks for \thCO{}. This result indicates that the gas temperature and/or density, and with it the pressure, is decreasing at these radii for over half of the sample. 

\section{Discussion} \label{sec:discussion}
In this section, we discuss the accuracy and precision of our results and their implications. First, we investigate the robustness of the pressure variations and derive the midplane pressure derivative for a subset of our sample, a quantity of central importance for dust evolution. Second, we look at the uncertainties, systematics, and biases in determining azimuthal velocities from molecular line observations. Lastly, we will place our results into the bigger picture and compare our findings to the predictions from theory. 

\subsection{Derivation of midplane pressure derivative} \label{sec:disc_pmid}
In this subsection, we discuss the role of the pressure gradient in generating observed substructures in the dust continuum. Early theoretical studies have shown that the (midplane) gas pressure derivative determines the overall evolution of large dust grains in disks, as it sets their drift rate \citep{Whipple_1972, Weidenschilling_1977}. While the observed dust continuum is emitted close to the disk midplane, as millimeter dust pebbles quickly settle \citep{Dullemond_Dominik_2004, Villenave_ea_2020}, CO molecular lines usually trace heights of $z/R\approx0.2-0.3$ \citep[e.g.,][]{Law_ea_2021b, Law_ea_2023, Galloway_exoALMA}. 

Given the vertical and radial thermal structure, defined by $c_\mathrm{s}\,(R, z)$, as estimated by \cite{Galloway_exoALMA} for a subset of our sample, we can overcome this spatial discrepancy and infer the midplane pressure gradient from the pressure gradient in the surface layers.

Following Appendix \ref{app:press_deriv}, it can be shown that the slope of the midplane pressure profile is
\begin{equation}
\label{eq:dPdr_mid}
     \left. \frac{\partial \ln P}{\partial \ln R} \right|_{z = 0} =  \left.\frac{\partial \ln P}{\partial \ln R}\right|_{z=z_\mathrm{emit}}  -  \left. \frac{\partial \ln \chi}{\partial \ln R} \right|_{z=z_\mathrm{emit}},
\end{equation}
where $\chi (R,z)$ is given by
\begin{equation}
    \chi (R,z) =  \exp\left( - \int_{z=0}^{z} \frac{\Omega_{\mathrm{k, mid}}^2\,\tilde{z}}{c_\mathrm{s}^2\,(R,\tilde{z}) }   \left[\frac{1}{1+\tilde{z}^2/R^2}\right]^{3/2}\,\mathrm{d}\tilde{z}\right)
\end{equation}
which is the ratio between the pressure at the emitting height $z$ and the midplane. 

In addition to the rotation curve, the derivations of the midplane pressure derivative require knowledge of disk emission height, 2D temperature structure, and stellar mass. For the latter, we take the $M_\star$ estimates from \cite{Longarini_exoALMA} for the analysis presented in this section, since they simultaneously fit both CO rotation curves with a prescription for each term of Eq.\,\ref{eq:vrot}, which provides a more accurate estimate of $M_\star$ than the \verb|discminer| kinematic stellar masses. However, since this procedure requires knowledge of the 2D temperature structure of the disk, it can only be employed on a subset of our sample. Because of this limitation and for consistency, we remind the reader that in the rest of this paper, we have used the \verb|discminer| kinematic stellar masses for our \dvphi{} analysis across the whole sample. To be consistent, we now also incorporate \cite{Longarini_exoALMA} derived self-gravity term into the subtraction of the background velocity (nominator Eq.\,\ref{eq:dPdr}) for sources that have non-negligible disk masses ($M_\mathrm{disk}> 0.05M_\star$), which are AA\,Tau, DM\,Tau, HD\,34282, J1842, LkCa\,15, and SY\,Cha. It's crucial to consider self-gravity in this context because it slightly increases the anticipated background velocity, by a few percent of the Keplerian speed \citep{Longarini_exoALMA}. This shift causes \dvphi{} to move to lower values, which is important to factor in as we are focusing on the points where \dvphi{} and hence \dpmid{} intersect at zero.

We estimate the uncertainty of each parameter on \dpmid{} via a Monte Carlo error propagation. We draw a total of 1000 samples, assuming Gaussian errors for the emission height (Eq. \ref{eq:em_height}) and 2D temperature structure (Eq. \ref{eq:2D_temp}) within the central 68\,\% of the Markov Chain Monte Carlo (MCMC) posterior distribution of the model parameters \citep[][]{Izquierdo_exoALMA, Galloway_exoALMA}. However, the dominant uncertainty in the derivation of the midplane pressure profile is the stellar mass that determines the background Keplerian velocity. Even though only a few percent of uncertainty in $M_\star$ is enough to dominate over the combined uncertainty of the other parameters, it generally only changes \dvphi{} by a constant everywhere, therefore not altering the shape of the rotation curve. Still, some (small) variations in the shape of the rotation curve may result from systematic uncertainties in the multi-level fitting procedure. A rigorous treatment of all the uncertainties and covariances between them remains challenging. Applying such a comprehensive framework for error propagation to our data is beyond the scope of this paper \citep[see][for a detailed study]{Andrews_ea_2024}. In our results we neglect the combined bootstrapped uncertainties from the other parameters and only display the uncertainty of $M_\star$ on the pressure profiles.  We vary $M_\star\pm \delta M_\star$ with $\delta M_\star=3\%$, which is the precision currently achievable with the applied rotation curve fitting procedures \citep{Andrews_ea_2024, Veronesi_ea_2024}.

\begin{figure}[t]
    \centering
    \includegraphics[width=1\linewidth]{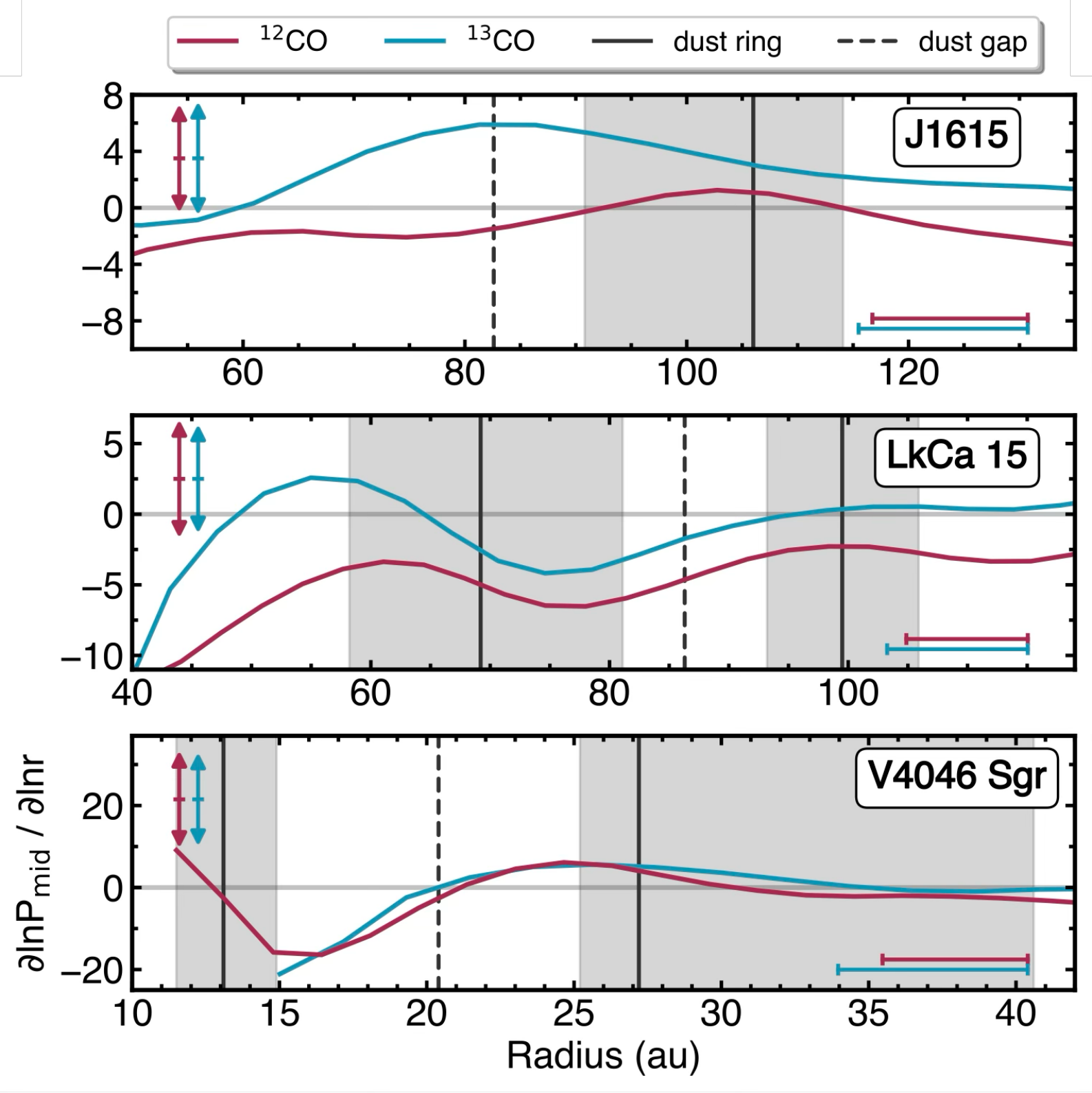}
    \caption{Radial profiles of \dpmid{} for selected sources, focused on the region of their continuum substructures. The vertical colored arrows in the upper left show variations in the stellar mass of $\pm$3\,\% which would result in a shift of the whole profiles up and down. We expect dust rings and gaps to be co-located with \dpmid{}=0 if induced by pressure maxima and minima (see Fig.~\ref{fig:vphi_press}~b). The vertical grey shaded areas show the width of the continuum rings \citep[][]{Curone_exoALMA}. The other plot annotations are the same as in Fig.~\ref{fig:dvphi_small}.}
    \label{fig:dlnPdlnR_selected}
\end{figure}

We present the profiles of the logarithmic radial derivative of the midplane pressure for the disks of J1615, LkCa\,15, and V4046\,Sgr in Figure~\ref{fig:dlnPdlnR_selected}. We chose to present these disks since their profiles show \dpmid{}=0 spatially coinciding with the location of the dust substructures within their uncertainties. However, this is not the case for all the disks and dust substructures, which will be discussed in the next subsection. In the Appendix, we present a gallery for the whole sub-sample in Fig.~\ref{fig:pressure_small} \& \ref{fig:pressure_large}.

We also made use of the 2D temperature structure and Equations~\ref{eq:dPdr}~and~\ref{eq:dPdr_mid} to assess the influence of the vertical temperature and pressure gradient on the slopes of our \dvphidr{}-profiles (Sec.~\ref{sec:res_p_variations}) when projecting \dvphi{} to the midplane. The slope did not change significantly within the given uncertainties, hence we can state that the analysis of pressure substructures using \dvphidr{} at the emitting height is robust to infer the underlying pressure substructures.

\subsection{Origins of pressure variations} \label{sec:disc_theory}
Until now, we remained agnostic on the origins of the substructures in \dvphi{} caused by underlying variations in pressure. In Section \ref{subsec:dvphi_inner_disk}, we report that the majority of dust rings and gaps are co-located with corresponding maxima or minima in the gas pressure traced by \dvphidr{}. 
The question remains whether those pressure substructures are mainly driven by changes in the local gas temperature or density. In the latter case, one can imagine gas gaps that are carved out by planets \citep[e.g.,][]{Kanagawa_ea_2015, Teague_ea_2018b} or gas density substructures caused by other mechanisms like self-induced dust traps, dead-zones, the vertical shear instability, or magnetically driven winds \citep[][for a review]{Bae_PPVII, Lesur_PPVII}. On the other hand, it has also been shown that strong radial gradients in temperature can lead to deviations from Keplerian rotation \citep{Keppler_ea_2019, Rab_ea_2020}. 

We can disentangle the two contributions to the pressure, by comparing the positive \dvphi{} gradients with the line width and peak brightness temperature radial profiles at these locations. In the presence of gas density gaps, we expect a local line width minimum compared to the unperturbed background emission, when traced by the optically thick CO lines. Their broadened line widths are dominated by the density of the species rather than the temperature or turbulence, which is the case for optically thin emission \citep{Hacar_ea_2016, Izquierdo_ea_2021}. Hence, when simultaneously detecting a positive \dvphi{} gradient and a minimum in the line width with a high (non-decreasing) gas temperature, \cite{Izquierdo_ea_2023} shows that the deviation from Keplerian rotation is the result of a surface density dip. Conversely, if the increased temperature of the gas gap would be the driving mechanism, \dvphi{} would be expected to show a negative radial gradient as has been studied by \cite{Rab_ea_2020}.

In Figures \ref{fig:delta_linewidth} and \ref{fig:peakint} of the Appendix, we show the line width residuals and peak intensity profiles for both CO molecules and all disks focused on the region of the continuum emission. Out of 17 continuum gaps with a positive CO \dvphi{} radial gradient, we identify nine that show a minimum in line widths, together with a warm gas temperature in either both or one of the CO molecules. This supports the interpretation of these pressure gaps caused by surface density minima. In column 8 of Table\,\ref{tab:substructures}, we highlight the gap substructures identified to be gas surface density gaps. We further give a rough estimate of planet masses potentially driving the \dvphi{} gap substructures in Section\,\ref{app:planet_mass} of the Appendix, using the relations derived in \cite{Yun_ea_2019}.

We can also see clear evidence of gas density cavities in the disks CQ Tau, J1604, J1852, and MWC\,758, as in those inner regions \dvphidr{} is positive, coinciding with a line width decrement and an increased peak intensity. Conversely, if the pressure gradient were dominated by strong changes in the disk temperature, we would expect \dvphidr{} to be decreasing \citep[e.g.,][]{Rab_ea_2020}. The \thCO{} rotation curves are expected to be more trustworthy in this analysis since their emission originates closer to the midplane and is more sensitive to variations in density rather than temperature, which is the case for \twCO{}. A good example is J1604, where \dvphi{} is radially decreasing in \twCO{}, though increasing for \thCO{}.  

In the case of dust continuum rings, we observe that approximately 75\% of them are located alongside pressure bumps. However, the remaining rings do not exhibit a noticeable increase (or even a decrease) in the radial gradient of \dvphi{} or \dpmid{} at their continuum peak location. This suggests that the underlying pressure structure may not be the exclusive cause of these rings. As discussed in Sec.~\ref{sec:discus_measure}, there might be shortcomings in the accurate measurement of \vphi{} at these close-in radial locations due to limited angular resolution. Some studies also show that dust rings can be produced without an underlying (static) pressure bump. Examples include ice lines \citep[e.g.,][]{Saito_Sirono_2011, Zhang_ea_2015, Hyodo_ea_2019}, second-generation dust rings caused by the collisions of planetesimal rings \citep[e.g.,][]{Jiang_Ormel_2021, Testi_ea_2022}, or transient features in the gas surface density such as zonal flows \citep[e.g.,][]{Uribe_ea_2011, Flock_ea_2015, Cui_Bai2021, Hsu_ea_2024}. In the latter, magneto-hydrodynamical simulations exhibit dust rings that no longer coincide with a pressure maximum that generated it, given a slow diffusion timescale of the rings in the case of low levels of turbulence \citep[][]{Riols_Lesur_2019}.  

The formation of continuum substructures is determined by the dynamics of the dust, which depends on the particle Stokes number and gas density. Consequently, one can also create dust rings through dust traffic jams \citep[e.g.,][]{Dullemond_Penzlin_2018}, coagulation front of pebbles \citep[e.g.,][]{Ohashi_ea_2021}, or the dust back-reaction on the gas \citep[e.g.,][]{Jiang_Ormel_2023}. The morphology of substructures can also depend on the grain size distribution, which will be investigated in a forthcoming paper by Winter et al. (in prep.). 

Lastly, the detection of kinematical substructures, potentially driven by variations in gas pressure, extending well beyond the dust continuum emission is intriguing (see Fig.\,\ref{fig:dvphi_large}). It is often suggested that the outermost gas pressure substructure(s) set the outer disk dust radius \citep[][]{Pinilla_ea_2020, Zormpas_ea_2022}, raising questions on the dust trapping efficiency of the gas substructures seen beyond $R_{\rm dust,95}$ or/and the detectability of dust trapped therein. Low dust trapping efficiency for mm dust particles can happen if the \textit{gas} surface density in the outermost regions is low ($St \gg 1, St \propto \Sigma_{\rm gas}^{-1}$), possibly resulting in low surface brightness, below detection limits. Dust diffusion is another possible explanation. This occurs for high values of $\alpha/St$, such that the dust traps are leaky and cannot retain small $\micron$-sized grains \citep[e.g. for $St \ll 1$ and high $\alpha$, see,][]{Dullemond_ea_2018, Rosotti_ea_2020}. Additionally, the global pressure gradient may still be negative (as seen for most disks in Fig.\,\ref{fig:pressure_large}) such that the pressure substructures do not act as dust traps, but only slow down the particle drift. Another possibility is that these pressure variations are short-lived and that those detected in the outer disk would have appeared after all the dust had drifted toward the inner disk. However, dynamic pressure bumps that appear and disappear over finite lifetimes — similar to zonal flows — are ineffective at preventing dust drift/loss over million-year timescales \citep[e.g.,][]{Stadler_ea_2022}.

\subsection{How well can we measure the rotational velocity?} \label{sec:discus_measure}
The measurement of the gas velocity in protoplanetary disks requires knowledge of the geometry of the disk, such as its center, position angle, inclination, and emission height of the molecular line. Once these parameters are retrieved via fitting of kinematical data, it is possible to project a grid onto the line-of-sight velocity centroid map to determine the azimuthal velocity within each projected radial annulus. The procedure assumes that the azimuthal velocity is axisymmetric along the disk minor axis and that it dominates over radial and vertical motions in the line-of-sight velocity (Eq.~\ref{eq:v_los}). The latter statement safely holds for intermediate and high inclinations (i.e., $i\approx$\,30-60\degree), as the Keplerian rotation is approximately ${\upsilon_{\rm kep}\approx 3.0\,\mathrm{km/s}\sqrt{M_\star/M_\odot}\sqrt{100\,\mathrm{au}/R}}$. Meanwhile, the sound speed only reaches $c_s\approx300\,\mathrm{m/s} \sqrt{T_\mathrm{gas}/25\,\mathrm{K}}$, providing an estimate of the magnitude of vertical and radial motions in the disk.

Still, it is possible to misinterpret velocity components due to problems in their decomposition. In particular, mistaking radial as azimuthal velocities and vice versa, due to non-axisymmetric kinematical features. The most obvious case would be for a protoplanet embedded in a disk that excite spiral density waves launched at the Lindblad resonances \citep[][]{Goldreich_Tremaine_1979, Goldreich_Tremaine_1980, Goodman_Rafikov_2001, Rafikov_2002} with associated non-axisymmetric (radial) velocity perturbations \citep{Bollati_ea_2021, Fasano_ea_2024}. In Appendix\,\ref{app:vel_ext_pl}, we present a parameter study on the extraction of \vphi{} from a hydrodynamical simulation with an embedded planet. We show that the misinterpretation of l.o.s.-components further depends on the azimuthal location of the planet in the disk as can be seen in Fig.~\ref{fig:planet_azimuth}. It is the strongest if the planet is located right in between the minor and major axes of the disk, and the deviations from the \textit{true} \dvphi{} velocity can be as big as 25\,\%. This is as expected, since at these azimuths both \vphi{} and \vrad{} contribute strongly to the $\upsilon_\mathrm{l.o.s.}$ (Eq.\,\ref{eq:vphi_los}+\ref{eq:vrad_los}). Reassuringly, the shape of the \dvphi{}-profile is hardly affected by this projection effect, hence the location of the inferred velocity and pressure substructures remain the same. 

Another uncertainty in the extraction of azimuthal velocities lies in the retrieval of the stellar mass and disk inclination \citep[][]{Andrews_ea_2024}. The inclination is important for the extraction of \vphi{} itself, while the stellar mass only gets introduced in \dvphi{}. For face-on disks, it can be hard to break the degeneracy of $M_\star \cdot \sin{i}$, which sometimes requires fixing the inclination to the value of the continuum. Nevertheless, for mid and high inclinations, the parameters can be robustly disentangled, and compare well to stellar masses reported in the literature \citep{Longarini_exoALMA}. 

Lastly, there can be biases in determining the rotational velocity through a shift of the line centroid within one beam introduced by strong gradients in intensity. Such gradients can be of either a physical nature, like deep gas cavities in transition disks, or due to insufficient angular resolution. In the latter case, we expect to retrieve sub-Keplerian motion if the intensity gradient becomes too steep within one beam, most apparent in the inner regions of disks (either radially increasing or decreasing), so the centroid is shifted towards higher intensities \citep{Keppler_ea_2019, Boehler_ea_2021, Andrews_ea_2024}. The same intensity shift occurs in the outermost regions of the disk, where the intensity diminishes and the signal-to-noise ratio (SNR) is low. In these areas, the CLEAN algorithm has difficulty capturing low-intensity diffuse emissions. The resulting noisy images can bias the estimates of location and velocity, causing the centroid to shift radially inward and leading to the retrieval of super-Keplerian rotation curves. Continuum subtraction (as applied on the cubes used in this work) can also affect the slope of the intensity profiles but has shown not to affect the retrieval of rotational velocities significantly \citep[e.g.,][]{Teague_ea_2018b, Izquierdo_ea_2023}. 

\vspace{4mm}
\subsection{Velocity offsets between molecular lines} \label{sec:disc_offset}
In this last subsection, we address the offset of \dvphi{}-substructures as traced by \twCO{} and \thCO{}. We expect disks to be vertically stratified, i.e. they show a warmer upper disk layer and get cooler towards the midplane, as seen in observations \citep{Law_ea_2021b, Galloway_exoALMA}. Therefore, the thermal pressure gradient is different at distinct heights in the disk, which leads to the differences of rotation between \twCO{} and \thCO{} shown in Sec.~\ref{subsec:vert_strat}.  Yet, there appears to be a dependency with radius on the level of vertical velocity stratification traced via the differences in rotation of \twCO{} and \thCO{}. Vertical stratification appears more pronounced at larger radii than smaller ones, thus radially more extended sources show a higher degree of stratification than smaller ones. The exceptions are the highly dynamically perturbed and small ($r<$150\,au) stratified disks of CQ\,Tau and MWC\,758, and the extended but very smooth non-stratified disk of V4046.

We see that particularly at large radii, \twCO{} and \thCO{} exhibit big differences in emission heights. In contrast, at short separations, the emission surfaces are located close to each other \citep[see Fig.\,5 in][]{Galloway_exoALMA}. In other words, \twCO{} and \thCO{} trace very distinct disk regions in temperature and density at large orbital radii, leading to substantial differences in their pressure gradient and rotation. This effect has recently been studied in detail by \cite{Pezzotta_ea_2025} and explains the vertical stratification observed at large distances for extended sources. It also clarifies why V4046 lacks any vertical stratification, as here the \twCO{} and \thCO{} emission surfaces do trace similar heights and thus pressure gradients throughout the disk \citep[][]{Izquierdo_exoALMA, Galloway_exoALMA}. The different vertical and radial temperature gradients between the CO molecules can also help to explain their different gradients in \dvphi{}. It has been shown that strong radial temperature gradients can lead to beam-smearing effects and thus artificial \dvphi{} variations \citep[][]{Keppler_ea_2019, Bohn2022, Pezzotta_ea_2025}. For a few sources, we observe opposing \dvphi{} gradients between \twCO{} and \thCO{} extending no more than 100 au in radius in the outer disk regions (see AA\,Tau, DM\,Tau, J1615 in Fig.~\ref{fig:dvphi_large_kms}). The peak intensity radial profiles show a \textit{negative} temperature gradient at those radii in both lines \citep[see Fig.~3 in][]{Galloway_exoALMA} which should result in a \textit{positive} \dvphi{} gradient if driven by the temperature variation in both lines. This is observed in \twCO{} for AA\,Tau \twCO{} ($R\sim$200-300\,au) and J1615 ($R\sim$400-480\,au), however, not in their \thCO{} curves.

Another possibility for the \dvphi{} mismatch is that not all disks are in centrifugal balance or vertical hydrostatic equilibrium in their respective layers, i.e. there are non-negligible asymmetric vertical and radial velocity flows, which can introduce mismeasurement of \vphi{} between the lines at those regions of interest. The predominantly negative midplane pressure derivatives (see Figure \ref{fig:pressure_small}) indicate steep pressure profiles, in particular for the transition disks of HD\,34282 and SY\,Cha, which are indicative of a centrifugal imbalance. Potentially, there could be fast (several 100's m/s) radial flows at these cavity edges \citep[e.g.,][]{Calcino_ea_2024}, hence the gas disk expands outwards, thus not being in centrifugal equilibrium. Another possibility is that such steep pressure profiles are signposts of the Rossby-Wave instability that has been shown to produce such features and vortices \citep[e.g.,][]{Meheut_ea_2010, Chang_ea_2023}. This is potentially traced in the three most asymmetric continuum disks in our sample, HD\,135344B, HD\,143006, and MWC\,758, which all exhibit continuum crescents \citep[][]{Curone_exoALMA, Wölfer_exoALMA}. Due to incomplete determination of the vertical physical structure, we could not derive their \dpmid{}-profiles, though their \dvphi{} profiles show negative gradients co-located with their dust continuum crescents indicative of pressure bumps (see Fig.~\ref{fig:dvphi_small}).

\section{Conclusions} \label{sec:concl}
In this sixth paper of the exoALMA paper series, we presented the rotation curves and deviations from Keplerian rotation (\dvphi{}) of the \twCOfull{}{} and  \thCOfull{} line emission for the entire exoALMA sample. 

\begin{enumerate}[i)]
    \item We show that the CO rotational velocities show clear evidence of vertical stratification, i.e. the \thCO{} molecular line rotates faster than the \twCO{}. This mismatch in rotation cannot be explained by differences in gravity and a constant temperature alone, indicating that the thermal pressure gradient differs at distinct heights in the disk, which is expected for disks with a hot surface layer and a cold midplane.
    
    \item  Substructures in \dvphi{} are ubiquitous in our sample, both on small ($\sim\!10$\,au) and large radial scales ($\sim\!100$\,au), indicative of strong pressure variations driving them. We also observe pressure substructures well beyond the dust continuum emission, superimposed on the overall steep pressure drop at the farthest radii. This raises the question of why the dust is not trapped by these substructures in the outer disk.

    \item We observe that a majority of continuum rings (75\%) and gaps (80\%) are co-located with gas pressure maxima and minima, respectively, as traced by the radial derivative of \dvphi{} of the CO molecular lines. We emphasize that these kinematic signatures are not manifested in the line \textit{intensities}, solely in the line centroid. Hence, we conclude that variations in gas pressure are likely the dominant mechanism for ring and gap formation in the dust continuum.

    \item Positive \dvphi{} radial gradients indicate variation in the gas density rather than temperature, indicating the presence of gas surface density gaps co-located with continuum gaps, potentially carved out by embedded protoplanets. 

    \item On large scales, we find most disks to rotate at sub-Keplerian speeds due to a negative pressure gradient induced by the drop in gas density and temperature at these outer radii. This has profound impacts on dust dynamics since it increases the radial inward drift of pebbles in the outer disk regions. This could rapidly replenish and enrich the inner disk with dust from the outer domains.
    
    \item For the first time, we determine the midplane pressure derivative from observational data. For a few sources, we find that \dpmid{}=0 is co-located with continuum substructures confirming the presence of underlying gas pressure bumps as predicted by theory \citep[e.g.,][]{Weidenschilling_1977, Birnstiel_ea_2010}.

    \item The data presented in this paper is publicly released as a Value-Added-Data-Product (VADP). The VADP release includes the radial profiles of rotation curves \vphi{}, the Keplerian background rotation \vkep{}, deviations from Keplerian rotation \dvphi{}, and the midplane pressure derivative for all three sets of continuum-subtracted image cubes. The properties of the reported \dvphi{} substructures listed in Table\,\ref{tab:substructures} will also be included in the release.
\end{enumerate}

In conclusion, this paper demonstrates that the study of rotational velocities traced via high-angular resolution molecular line ALMA observations opens a unique window to comprehensively analyze the underlying dynamical and pressure substructures of planet-forming disks. Future work investigating the properties and abundance of these substructures will help us to further unravel their physical nature, bringing us a step closer in our pursuit to detect protoplanets in formation.

\bibliography{mybib}{}
\bibliographystyle{aasjournal}
\section*{Acknowledgments}
The authors would like to thank the anonymous referee for the thoughtful and constructive feedback, which significantly improved the quality of this work, as well as Haochang Jiang, Bin Ren, and Richard Booth for their helpful discussions.
This paper makes use of the following ALMA data: ADS/JAO.ALMA\#2021.1.01123.L. ALMA is a partnership of ESO (representing its member states), NSF (USA) and NINS (Japan), together with NRC (Canada), MOST and ASIAA (Taiwan), and KASI (Republic of Korea), in cooperation with the Republic of Chile. The Joint ALMA Observatory is operated by ESO, AUI/NRAO and NAOJ. The National Radio Astronomy Observatory is a facility of the National Science Foundation operated under cooperative agreement by Associated Universities, Inc. We thank the North American ALMA Science Center (NAASC) for their generous support including providing computing facilities and financial support for student attendance at workshops and publications.

JS, MB, and DF have received funding from the European Research Council (ERC) under the European Union’s Horizon 2020 research and innovation programme (PROTOPLANETS, grant agreement No. 101002188). JS has performed computations on the `Mesocentre SIGAMM' machine, hosted by Observatoire de la Cote d’Azur. AJW has received funding from the European Union’s Horizon 2020 research and innovation programme under the Marie Skłodowska-Curie grant agreement No 101104656. Support for AFI was provided by NASA through the NASA Hubble Fellowship grant No. HST-HF2-51532.001-A awarded by the Space Telescope Science Institute, which is operated by the Association of Universities for Research in Astronomy, Inc., for NASA, under contract NAS5-26555. CL has received funding from the European Union's Horizon 2020 research and innovation program under the Marie Sklodowska-Curie grant agreement No. 823823 (DUSTBUSTERS) and by the UK Science and Technology research Council (STFC) via the consolidated grant ST/W000997/1. PC acknowledges support by the Italian Ministero dell'Istruzione, Universit\`a e Ricerca through the grant Progetti Premiali 2012 – iALMA (CUP C52I13000140001) and by the ANID BASAL project FB210003.
JB acknowledges support from NASA XRP grant No. 80NSSC23K1312. NC has received funding from the European Research Council (ERC) under the European Union Horizon Europe research and innovation program (grant agreement No. 101042275, project Stellar-MADE). SF is funded by the European Union (ERC, UNVEIL, 101076613), and acknowledges financial contribution from PRIN-MUR 2022YP5ACE. MF is supported by a Grant-in-Aid from the Japan Society for the Promotion of Science (KAKENHI: No. JP22H01274). CH acknowledges support from NSF AAG grant No. 2407679. IH acknowledges a Research Training Program scholarship from the Australian Government. TH is supported by an Australian Government Research Training Program (RTP) Scholarship. JDI acknowledges support from an STFC Ernest Rutherford Fellowship (ST/W004119/1) and a University Academic Fellowship from the University of Leeds. GL has received funding from the European Union’s Horizon 2020 research and innovation program under the Marie Sklodowska-Curie grant agreement No. 823823 (DUSTBUSTERS). FMe has received funding from the European Research Council (ERC) under the European Union's Horizon Europe research and innovation program (grant agreement No. 101053020, project Dust2Planets). CP acknowledges Australian Research Council funding via FT170100040, DP18010423, DP220103767, and DP240103290. DP acknowledges Australian Research Council funding via DP18010423, DP220103767, and DP240103290. GR acknowledges funding from the Fondazione Cariplo, grant no. 2022-1217, and the European Research Council (ERC) under the European Union’s Horizon Europe Research \& Innovation Programme under grant agreement no. 101039651 (DiscEvol). H-WY acknowledges support from National Science and Technology Council (NSTC) in Taiwan through grant NSTC 113-2112-M-001-035- and from the Academia Sinica Career Development Award (AS-CDA-111-M03). GWF acknowledges support from the European Research Council (ERC) under the European Union Horizon 2020 research and innovation program (Grant agreement no. 815559 (MHDiscs)). GWF was granted access to the HPC resources of IDRIS under the allocation A0120402231 made by GENCI. TCY acknowledges support by Grant-in-Aid for JSPS Fellows JP23KJ1008. Support for BZ was provided by The Brinson Foundation. Views and opinions expressed by ERC-funded scientists are however those of the author(s) only and do not necessarily reflect those of the European Union or the European Research Council. Neither the European Union nor the granting authority can be held responsible for them. 

\software{astropy \citep{Astropy_2022}, CARTA \citep{Comrie_2021_carta}, casa \citep{casa},  Discminer \citep{Izquierdo_ea_2021}, Eddy \citep{Teague_2019_eddy}, Grammarly (last used 01/2025, \href{https://www.grammarly.com/grammar}{website}), Matplotlib \citep{Hunter_mpl}, Numpy \citep{vanderWalt_np}, Scipy \citep{Virtanen_scipy} }

\newpage
\appendix
\counterwithin{figure}{section}

\section{Substructures} \label{app:substructures}
\startlongtable
\begin{deluxetable*}{lcccccc|c|l}
\tabletypesize{\scriptsize}
\tablewidth{1\textwidth} 
\tablecaption{Properties of the \dvphi{} substructures co-located with continuum gaps and rings}
\tablehead{
\colhead{Source} & \colhead{Feature} & \colhead{Tracer}    &   \colhead{Radial location}   &   \colhead{Width}   &  \colhead{Amplitude} &  \colhead{Offset} &  \colhead{$\Sigma_\mathrm{gap}$} & \colhead{Comments} \\
\colhead{}  &  \colhead{}    & \colhead{}    &  \colhead{(au, mas)}   &  \colhead{(au, mas)}   & \colhead{(m/s, \%~\vkep{}, M$_{\rm jup}$)} & \colhead{(au, m/s)}}
\colnumbers
\startdata 
AA\,Tau   &  B42 & \twCO{} & 47$\pm$5, 350$\pm$37 &  20$\pm$5, 150$\pm$37  & 314$\pm$116, 9.1 &  5, 107 & \nodata & \nodata \\
          &      & \thCO{} & 44$\pm$6, 330$\pm$45 &  8$\pm$6, 60$\pm$45  & 237$\pm$105, 6.3 &  2, -15 & \nodata & width $<$ one beamsize \\
         &  D64 & \twCO{} & 64$\pm$5, 475$\pm$37 &  13$\pm$5, 100$\pm$37  & 135$\pm$57, 4.5, \textbf{0.4} &  -0, 18 & \checkmark & \nodata \\
         &      & \thCO{} &  63$\pm$6, 465$\pm$45 &  12$\pm$6, 90$\pm$45  & 107$\pm$43, 3.4 &  -2, -34 & \checkmark & \nodata  \\
         &  B72 & \twCO{} & 74$\pm$5, 550$\pm$37 &  7$\pm$5, 50$\pm$37  & 99$\pm$47, 3.6 &  2, 36 & \nodata & width $<$ one beamsize \\
         &     & \thCO{} & 73$\pm$6, 540$\pm$45 &  8$\pm$6, 60$\pm$45  & 61$\pm$50, 2.1 &  1, -11 & \nodata & width $<$ one beamsize \\
         &  D80   & \thCO{} &  81$\pm$6, 600$\pm$45 &  8$\pm$6, 60$\pm$45  & 95$\pm$56, 3.4, \textbf{0.3} &  1, 6 & O & width $<$ one beamsize \\
         &  B90 & \twCO{} & 91$\pm$5, 675$\pm$37 &  7$\pm$5, 50$\pm$37  & 33$\pm$44, 1.3 &  1, -20 & \nodata & width $<$ one beamsize \\
         &      & \thCO{} & 95$\pm$6, 705$\pm$45 &  20$\pm$6, 150$\pm$45  & 81$\pm$46, 3.1 &  5, 13 & \nodata & center outside of ring width \\
\hline
DM\,Tau   &  D72 & \twCO{} & 73$\pm$4, 506$\pm$29 &  14$\pm$4, 95$\pm$29  & 213$\pm$40, 9.4, \textbf{1.2} &  1, -25 & O & \nodata \\
         &  B90 & \twCO{} & 94$\pm$4, 649$\pm$29 &  11$\pm$4, 76$\pm$29  & 78$\pm$30, 3.9 &  4, 37 & \nodata & \nodata \\
         &      & \thCO{} & 91$\pm$4, 630$\pm$29 &  17$\pm$4, 120$\pm$29  & 96$\pm$24, 4.5 &  1, -39 & \nodata & \nodata \\
\hline
HD\,135344B& B51 & \twCO{} & 43$\pm$8, 319$\pm$56 &  15$\pm$8, 112$\pm$56  & 215$\pm$92, 3.8 &  -8, 189 & \nodata & \nodata \\
         &      & \thCO{} & 43$\pm$8, 319$\pm$56 &  15$\pm$8, 112$\pm$56  & 154$\pm$86, 2.6 &  -8, 152 & \nodata & \nodata \\
         & D66 & \twCO{} & 61$\pm$8, 450$\pm$56 &  20$\pm$8, 150$\pm$56  & 157$\pm$99, 3.3 &  -6, 160 & O & high amplitude uncertainty \\
         &     & \thCO{} & 61$\pm$8, 450$\pm$56 &  20$\pm$8, 150$\pm$56  & 58$\pm$79, 1.2 &  -6, 105 &  O & high amplitude uncertainty \\
\hline
HD\,143006 &  B40 & \twCO{}& 38$\pm$6, 225$\pm$37 &  17$\pm$6, 100$\pm$37  & 657$\pm$189, 11.7 &  -3, 488 & \nodata & \nodata \\
         &  B64  & \twCO{} & 65$\pm$6, 387$\pm$37 &  21$\pm$6, 125$\pm$37  & 136$\pm$48, 3.2 &  0, 123 & \nodata & \nodata \\
         &       & \thCO{} & 72$\pm$9, 431$\pm$56 &  19$\pm$9, 112$\pm$56  & 170$\pm$42, 4.1 &  8, 53 & \nodata& center outside of ring width \\
\hline
HD\,34282 & D59 & \twCO{} & 90$\pm$8, 291$\pm$27 &  67$\pm$8, 218$\pm$27  & 788$\pm$123, 21.5 &  31, -97 &\checkmark & \nodata \\
         &     & \thCO{} & 85$\pm$12, 275$\pm$37 &  46$\pm$12, 150$\pm$37  & 609$\pm$271, 15.4 &  26, -104 & O & \nodata \\
         & B124& \twCO{} & 129$\pm$8, 418$\pm$27 &  11$\pm$8, 36$\pm$27  & 293$\pm$56, 9.7 &  5, 150 & \nodata & \nodata \\
         &     & \thCO{} & 139$\pm$12, 450$\pm$37 &  62$\pm$12, 200$\pm$37  & 202$\pm$83, 6.5 &  15, 100 & \nodata & \nodata \\
\hline
J1604    & B82 & \twCO{} & 97$\pm$6, 669$\pm$41 &  36$\pm$6, 246$\pm$41  & 279$\pm$62, 8.0 &  15, 35 & \nodata & center outside of ring width  \\
         &     & \thCO{} & 89$\pm$8, 619$\pm$56 &  49$\pm$8, 338$\pm$56  & 286$\pm$45, 7.8 &  7, 46 & \nodata & center outside of ring width \\
\hline
J1615   & D83  & \twCO{}& 93$\pm$7, 599$\pm$45 &  37$\pm$7, 239$\pm$45  & 89$\pm$28, 2.8, \textbf{0.6} &  11, 95 & \checkmark & center outside of gap width \\
        &      & \thCO{}& 66$\pm$8, 423$\pm$49 &  30$\pm$8, 195$\pm$49  & 200$\pm$48, 5.1, \textbf{0.9}  &  -17, -13 & \checkmark & center outside of gap width \\
        &  B106 & \thCO{}& 99$\pm$8, 635$\pm$49 &  36$\pm$8, 228$\pm$49  & 91$\pm$29, 2.9 &  -7, 41 & \nodata & \nodata \\
\hline
J1852   & D31 & \twCO{} & 36$\pm$6, 248$\pm$39 &  19$\pm$6, 130$\pm$39  & 191$\pm$137, 3.9, \textbf{0.6} &  5, 196 & \checkmark & high amplitude uncertainty \\
         & B50 & \twCO{} & 54$\pm$6, 365$\pm$39 &  15$\pm$6, 104$\pm$39  & 310$\pm$29, 7.7 &  4, 136 & \nodata & \nodata \\
         &     & \thCO{} & 58$\pm$7, 393$\pm$45 &  18$\pm$7, 121$\pm$45  & 129$\pm$33, 3.3 &  8, 33 & \nodata & center outside of ring width \\
\hline
LkCa\,15  & B68 & \twCO{} & 70$\pm$5, 442$\pm$32 &  10$\pm$5, 65$\pm$32  & 137$\pm$51, 3.7 &  1, 16 & \nodata & \nodata \\
         &     & \thCO{} & 69$\pm$6, 437$\pm$37 &  12$\pm$6, 75$\pm$37  & 237$\pm$36, 6.3 &  1, -1 & \nodata & \nodata \\
         & D86 & \twCO{}& 88$\pm$5, 561$\pm$32 &  27$\pm$5, 173$\pm$32  & 136$\pm$38, 4.2, \textbf{0.7} &  2, 15 & O & \nodata \\
         &     & \thCO{} & 82$\pm$6, 525$\pm$37 &  16$\pm$6, 100$\pm$37  & 142$\pm$33, 4.1, \textbf{0.7} &  -4, -49 & \checkmark & \nodata \\
        & B100 &\twCO{} & 109$\pm$5, 690$\pm$32 &  14$\pm$5, 86$\pm$32  & 87$\pm$22, 3.0 &  9, 39 & \nodata & center outside of ring width \\
\hline
MWC\,758   & D32 & \twCO{}  & 37$\pm$6, 237$\pm$40 &  16$\pm$6, 105$\pm$40  & 849$\pm$105, 15.1 &  5, 327 & \checkmark & \nodata \\
        &      & \thCO{}  & 44$\pm$7, 279$\pm$47 &  19$\pm$7, 124$\pm$47  & 687$\pm$142, 12.6 &  11, 4 & (\checkmark) & cold density gap \\
         & B51 & \twCO{} & 60$\pm$6, 382$\pm$40 &  29$\pm$6, 185$\pm$40  & 836$\pm$136, 18.9 &  9, 333 &\nodata & center outside of ring width \\
         &     & \thCO{} & 68$\pm$7, 434$\pm$47 &  29$\pm$7, 186$\pm$47  & 529$\pm$92, 12.1 &  17, 83 & \nodata & center outside of ring width \\
\hline
SY\,Cha   & B101 & \twCO{} & 116$\pm$10, 638$\pm$56 &  27$\pm$10, 150$\pm$56  & 119$\pm$49, 5.1 &  15, 123 & \nodata & \nodata \\
         &      & \thCO{} & 102$\pm$10, 562$\pm$56 &  27$\pm$10, 150$\pm$56  & 122$\pm$58, 4.5 &  1, -48 & \nodata & \nodata \\
\hline
V4046\,Sgr& B13  & \twCO{} & 14$\pm$2, 195$\pm$34 &  5$\pm$2, 69$\pm$34  & 566$\pm$251, 5.4 &  1, 43 & \nodata & \nodata \\
         &  D20 & \twCO{} &  21$\pm$2, 287$\pm$34 &  8$\pm$2, 115$\pm$34  & 431$\pm$125, 5.0, \textbf{0.7} &  0, -24 & \checkmark & \nodata \\
         &     & \thCO{} & 19$\pm$3, 270$\pm$45 &  9$\pm$3, 120$\pm$45  & 434$\pm$170, 4.8, \textbf{0.6} &  -1, 4 & O & \nodata \\
         &  B27 & \twCO{} & 30$\pm$2, 425$\pm$34 &  12$\pm$2, 161$\pm$34  & 185$\pm$53, 2.6 &  3, 99 & \nodata & \nodata \\
         &      & \thCO{} & 33$\pm$3, 465$\pm$45 &  6$\pm$3, 90$\pm$45  & 175$\pm$28, 2.6 &  6, 35 & \nodata & \nodata \\
\enddata
\tablecomments{Column~(1): target name. Column~(2): annular substructure label as in \cite{Curone_exoALMA}. “B" (for \textit{bright}) indicates a ring, while “D" (for \textit{dark}) indicates a gap. The number on the label is the feature distance from the central star measured in au. Column~(3): observational tracer. Column~(4-6): \dvphi{} substructure radial location, width, and amplitude inferred as explained in Sect.~\ref{sec:report_substr}. Column~(7): radial and velocity offset from the center of continuum substructure and \dvphi{}=0, respectively. Column~(8): detection (\checkmark) of a dip in line widths together with non-decreasing peak intensities at continuum gap location indicative of a co-located surface density gap $\Sigma_{\rm gap}$ as discussed in Sect.~\ref{sec:disc_theory}. No decrease in line width was detected for gap rows marked with ``O''. Column~(9): comments regarding irregularities. Comments on \dvphi{} substructures co-located with continuum substructures non-reported in this table can be found in Table~\ref{tab:non_substructures} in the Appendix. Radial \dvphi{}-profile with annotated substructures reported in this table can be found in Figures~\ref{fig:dvphi_substr_12co}~\&~\ref{fig:dvphi_substr_13co}.}

\label{tab:substructures}
\end{deluxetable*}


\newpage
\startlongtable
\begin{deluxetable*}{llc|l}
\tabletypesize{\scriptsize}
\tablewidth{1\textwidth} 
\tablecaption{Comments to non-reported \dvphi{} at continuum substructures}
\tablehead{
\colhead{Source} & \colhead{Feature} & \colhead{Tracer}  &  \colhead{Comments}}
\colnumbers
\startdata 
AA Tau    & D11 & \twCO{}, \thCO{}&  not accessible due to limited angular resolution \\
         & D80 & \twCO{}  & flat \dvphi{} profile \\
\hline
CQ Tau    & B41 & \twCO{}, \thCO{} & reversed and flat \dvphi{} gradient, respectively \\
\hline
DM Tau   & D14 & \twCO{}, \thCO{}  & not accessible due to limited angular resolution \\
         & B24 & \twCO{} & reversed (positive) \dvphi{} gradient \\
         & B24 & \thCO{} & not accessible due to limited angular resolution \\
         & D72 & \thCO{} & reversed (negative) \dvphi{} gradient \\
\hline
HD 135344B& D13 & \twCO{}, \thCO{}  & not accessible due to limited angular resolution \\
         & D66, B78 & \twCO{}, \thCO{} & asymmetric ring (vortex) with negative \dvphi{} gradient \\
\hline
HD 143006 & B7, D22 & \twCO{}, \thCO{}  & not accessible due to limited angular resolution \\
         & B40 & \thCO{} & reversed (positive) \dvphi{} gradient \\
         & D52 & \twCO{}, \thCO{} & flat and reversed \dvphi{} gradient, respectively \\
\hline
HD 34282 & D22 & \twCO{}, \thCO{}  & not accessible due to limited angular resolution \\
         & B47 & \twCO{}  & reversed (positive) \dvphi{} gradient \\
         & B47 & \thCO{}  & not accessible due to limited angular resolution \\
\hline
J1615  & D12, B26 & \twCO{}, \thCO{}  & not accessible due to limited angular resolution \\
       & B106 & \twCO{}   & reversed (positive) \dvphi{} gradient \\
\hline
J1842  & B36 & \twCO{}  & reversed (positive) \dvphi{} gradient \\
       & B36 & \thCO{}  & not accessible due to limited angular resolution \\
\hline
J1852  & B19 & \twCO{}, \thCO{}  & not accessible due to limited angular resolution \\
        & D31 & \thCO{}  & not accessible due to limited angular resolution \\
\hline
LkCa 15  & D15 & \twCO{}, \thCO{}  & not accessible due to limited angular resolution \\
         & B100 & \thCO{} & flat \dvphi{} profile \\
\hline
MWC758  & D64, B88 & \twCO{}, \thCO{}  & asymmetric ring (vortex) with negative \dvphi{} gradient  \\
\hline
SY Cha  & D33 & \twCO{}, \thCO{}  & not accessible due to limited angular resolution \\
\hline
V4046 Sgr& D8 & \twCO{}, \thCO{} & not accessible due to limited angular resolution \\
         & B13 & \thCO{} & not accessible due to limited angular resolution \\
\enddata
\tablecomments{Column~(1): target name. Column~(2): annular substructure label. “B" (for \textit{bright}) indicates a ring, while “D" (for \textit{dark}) indicates a gap. The number on the label is the feature distance from the central star measured in au. Column~(3): Observational tracer of \dvphi{}. Column~(4): Comments for why substructures were not reported. Note that dust substructures inaccessible due to limited angular resolution are not included in the total count of non-aligning \dvphi{}-substructures considered.}
\label{tab:non_substructures}
\end{deluxetable*}

\newpage
\section{Planet masses from \dvphi{} perturbations} 
\label{app:planet_mass}
In this subsection, we provide rough estimates of planet masses that are potentially driving the \dvphi{} perturbations observed co-located with the dust continuum gaps reported in Table~\ref{tab:substructures}. To this end, we follow the approach of \cite{Yun_ea_2019} that relates the gap width of the \dvphi{} perturbation to the planet mass via a set of 2D hydrodynamical simulations. We estimate the planet mass $M_p$ by numerically solving their equation 20
\begin{equation}
    \frac{W_\upsilon}{h_p} = 2.66 \left(\frac{M_p}{M_{th}}\right)^{-0.41}+ 2.04 \left(\frac{M_p}{M_{th}}\right)^{0.42},
\label{eq:Yun_20}
\end{equation}
where $W_\upsilon$ is the \dvphi{} gap-width, $h_p$ the pressure scaleheight at the planet's radial location $r_p$, and $M_{th}\equiv M_\star(h_p/r_p)^3$ the thermal mass. The pressure scale is given via $h_p=c_s \Omega_K$, where $\Omega_K$ is the Keplerian frequency. The soundspeed $c_s$ gets evaluated at the disk midplane using the midplane temperatures as measured in \cite{Galloway_exoALMA} (see Eq.\,\ref{eqn:tmid}). Disks for which no 2D temperature profile could have been measured, we estimate the midplane temperature via the expected irradiation temperature following \cite{Galloway_exoALMA}
\begin{equation}
\label{eqn:t_irr}
T_{\rm{irr}} (r) = \left(\frac{L_*}{8 \pi R^2 \sigma_{SB}} \right)^{1/4},
\end{equation}
where $L_*$ is the stellar luminosity, $R$ is the distance from the central star,  $\sigma_{SB}$ is the Stefan-Boltzmann constant \citep{DAlessio_ea_2001}. We report the rough planet mass estimates following the above equations in Table\,\ref{tab:planet_masses}.
\startlongtable
\begin{deluxetable*}{lcccccc}
\tabletypesize{\scriptsize}
\tablewidth{1\textwidth} 
\tablecaption{Planet masses derived from \dvphi{} gap properties}
\tablehead{
\colhead{Source} & \colhead{Feature} & \colhead{Tracer}    &   \colhead{Radial location}   &   \colhead{Width}   & \colhead{Pressure scaleheight} & \colhead{Planet mass} \\
\colhead{}  &  \colhead{}    & \colhead{}    &  \colhead{$r_p$: [au]}   &  \colhead{$W_\upsilon$: [au]}  &  \colhead{$h_p(r_p)$: [au]} & \colhead{[M$_{\rm jup}$]}}
\colnumbers
\startdata 
AA\,Tau  &  D64 & \twCO{} & 64$\pm$5  &  13$\pm$5 & 4.3 & 2.4  \\
         &      & \thCO{} &  63$\pm$6 &  12$\pm$6 & 4.2 &  2.6  \\
         &  D80 & \thCO{} &  81$\pm$6 &  8$\pm$6  & 5.9 & 0.44 \\
\hline
DM\,Tau   &  D72 & \twCO{} & 73$\pm$4  &  14$\pm$4   & 8.8 & 1.1  \\
\hline
HD\,135344B & D66 & \twCO{} & 61$\pm$8  &  20$\pm$8   & 6.0$^\star$ & 2.1 \\
            &     & \thCO{} & 61$\pm$8 &  20$\pm$8    &  6.0$^\star$ & 2.2 \\
\hline
HD\,34282 & D59 & \twCO{} & 90$\pm$8  &  67$\pm$8     &  7.7 & 1.4 \\
          &     & \thCO{} & 85$\pm$12  &  46$\pm$12   &  7.2 & 1.4 \\
\hline
J1615   & D83  & \twCO{}& 93$\pm$7 &  37$\pm$7  & 8.3 & 1.1 \\
        &      & \thCO{}& 66$\pm$8 &  30$\pm$8  & 5.2 & 2.5 \\
\hline
J1852   & D31 & \twCO{} & 36$\pm$6 &  19$\pm$6  &  2.9 & 16$^\dagger$  \\
\hline
LkCa\,15 & D86 & \twCO{}& 88$\pm$5  &  27$\pm$5  &  7.0 & 0.82\\
         &     & \thCO{} & 82$\pm$6 &  16$\pm$6 &  6.3 & 0.75 \\
\hline
MWC\,758   & D32 & \twCO{}  & 37$\pm$6  &  16$\pm$6   &  4.1$^\star$ & 2.7  \\
           &      & \thCO{}  & 44$\pm$7 &  19$\pm$7  &   5.0$^\star$ & 3.1 \\
\hline
V4046\,Sgr &  D20 & \twCO{} &  21$\pm$2 &  8$\pm$2   &  1.0 & 5.9 \\
           &     & \thCO{} & 19$\pm$3   &  9$\pm$3   &  0.9 & 5.7  \\
\enddata
\tablecomments{Column~(1): target name. Column~(2): annular substructure label as in \cite{Curone_exoALMA}, “D" (for \textit{dark}) indicates a gap. The number on the label is the dust feature's distance from the central star measured in au. Column~(3): observational tracer. Column~(4+5): \dvphi{} substructure radial location and width same as Tab.\,\ref{tab:substructures} inferred as explained in Sect.~\ref{sec:report_substr}. Column~(7): Gas pressure scaleheight at \dvphi{} radial location computed using gas midplane temperatures from \cite{Galloway_exoALMA}. Column~(8): Planet mass estimates in masses of Jupiter derived from the \dvphi{} width using Eq.\,\ref{eq:Yun_20} as in \cite{Yun_ea_2019}. Radial \dvphi{}-profile with annotated substructures reported in this table can be found in Figures~\ref{fig:dvphi_substr_12co}~\&~\ref{fig:dvphi_substr_13co}.}

\tablenotetext{^{\star}}{Gas pressure scale height computed via stellar luminosity midplane irradiation using Eq.\,\ref{eqn:t_irr} \citep[MWC\,758: $L_\star=20L_\odot$ and HD\,135344B $L_\star=5.1L_\odot$ from][, respectively]{Mannings_Sargent_2000, Guzman-Diaz_ea_2021}.}
\tablenotetext{^{\dagger}}{High planet mass estimate due to hot midplane temperature $T_{\rm mid}(r_p)=43$K \citep[see Fig.\,5 in][]{Galloway_exoALMA}.}
\label{tab:planet_masses}
\end{deluxetable*}


\section{Derivation of the radial gradient of the midplane pressure} \label{app:press_deriv}
To quantify the pressure variations in the midplane, we first assume that the local disc at the cylindrical radius $R$ is in vertical hydrostatic equilibrium such that:
\begin{equation}
\label{eq:Pz_ODE}
   \left. \frac{ \partial P}{\partial z}\right|_R  = \left. \frac{ \partial (\rho_\mathrm{gas} c_\mathrm{s}^2) }{\partial z}  \right|_R = 
 - \frac{G M_\star \rho_\mathrm{gas} z}{(R^2+z^2)^{3/2}}.
\end{equation}
Here, we have already imposed a functional form for $c_\mathrm{s}(R, z)$ such that the above equation can be directly integrated over $z$ to yield the expression:
\begin{equation}
    \rho_\mathrm{gas}(R,z) = \frac{\rho_0\,\chi}{g}
\end{equation}
where
\begin{equation}
    \chi (R,z) =  \exp\left( - \int_0^{z} \frac{\Omega_{\mathrm{kep, mid}}^2  \,\tilde{z}}{c_\mathrm{s}^2(\tilde{z}) }   \left[\frac{1}{1+\tilde{z}^2/R^2}\right]^{3/2}\,\mathrm{d}\tilde{z}\right)
\end{equation}
and
\begin{equation}
   g(R, z)  =  \frac{c_\mathrm{s}^2}{c_\mathrm{s,mid}^2} = \frac{T}{T_\mathrm{mid}}.
\end{equation} 
The radial pressure gradient at a given height $z$ is:
\begin{equation} \label{eq:dlnPdlnr}
    \left.\frac{\partial \ln P}{\partial \ln R}\right|_{z} =\frac{\partial \ln \rho_0}{\partial \ln R} + \frac{\partial \ln c_\mathrm{s,mid}^2}{\partial \ln R}  +  \left.\frac{\partial \ln \chi}{\partial \ln R}\right|_{z} .
\end{equation}
The first two components on the right hand side of equation~\ref{eq:dlnPdlnr} are:
\begin{equation}
    \partiallndr{\rho_0}  = \partiallndr{\Sigma_\mathrm{g}}  - \partiallndr{H}
\end{equation}where
\begin{equation}
    \frac{\partial \ln H}{\partial \ln R}  =  \frac{\partial \ln c_\mathrm{s, mid}}{\partial \ln R}+ \frac{3}{2} .
\end{equation} The last component of equation~\ref{eq:dlnPdlnr} can be written:
      \begin{equation}
    \label{eq:dlnchidlnR}
 \frac{\partial \ln  \chi} {\partial  \ln R}  =   -  \int_0^z \frac {f \tilde{z}}{gH^2} \left\{ \frac{\partial \ln  f} {\partial  \ln R} - \frac{\partial \ln  g} {\partial  \ln R}  -2  \frac{\partial \ln  H} {\partial  \ln R}   \right\} \,\mathrm{d} \tilde{z} 
      \end{equation}
    where we have defined:
     \begin{equation}
         f(R, z) =\left( \frac{1}{1+z^2/R^2}\right)^{3/2}.
     \end{equation}

So far, all of the components of equation~\ref{eq:dlnchidlnR} are general. Then, we assume a functional form for the temperature structure derived by \citet{Dartois_ea_2003} and fitted for in \cite{Galloway_exoALMA}, where the midplane temperature $T_\mathrm{atm}$ and atmospheric temperature $T_\mathrm{atm}$ are described by power-laws: 
\begin{equation} \label{eqn:tatm}
T_{\rm{atm}} (R) = T_{\rm{atm}, p} \left( R / 100\,\text{au}\right)^{-q_{\rm{atm}}}
\end{equation}
\begin{equation} \label{eqn:tmid}
T_{\rm{mid}} (R) = T_{\rm{mid}, p} \left( R / 100\,\text{au}\right)^{-q_{\rm{mid}}}.
\end{equation}
At each radius and height between the atmosphere and midplane, the temperature is then smoothly connected using a squared cosine function:
\begin{equation} \label{eq:2D_temp}
T(R, z)= \begin{cases}T_{\text {atm }}(R)+ [ \, T_{\text {mid }}(R)-T_{\text {atm }}(R) ]\, \cos ^2\left(\frac{\pi}{2} \frac{z}{Z_q}\right) & \text { if } z<Z_q \\ 
T_{\text {atm }}(R) & \text { if } z \geq Z_q , \end{cases}
\end{equation}
where $Z_q$ is also given by a power-law $Z_q(R)=Z_0(R/100\,\text{au})^{q_{\rm z}}$. Then the radial derivative of $g(R, z)=T/T_\mathrm{mid}$ is denoted by
\begin{equation}
    \partiallndr{g} = \partiallndr{T} - \partiallndr{T_\mathrm{mid}}.
\end{equation}
where 
\begin{equation}
\partiallndr{T} = \begin{cases} -\frac{1}{T}\left[q_{\text {atm }} \cdot T_{\text {atm }}\left(1+\left(\frac{q_{\text {mid }} T_{\text {mid }}}{q_{\text {atm }} T_{\text {atm }}}-1\right) \cdot \cos ^2\left(\frac{\pi}{2} \frac{z}{Z_q}\right)\right)
+ \frac{\pi}{2} \frac{q_z \cdot z}{Z_q}\left(T_{\text {atm }}-T_{\text {mid }}\right) \sin \left(\pi \frac{z}{Z_q}\right)\right] & \text { if } z<Z_q \\
- q_\mathrm{atm} & \text { if } z \geq Z_q ,\end{cases}
\end{equation}
and
\begin{equation}
    \partiallndr{T_\mathrm{mid}} = \frac{R}{T_\mathrm{mid}} \frac{\partial T_\mathrm{mid}}{\partial R} = - q_\mathrm{mid}.
\end{equation}

\vspace{0.5cm}
\section{Velocity extraction dependency on planet position} 
\label{app:vel_ext_pl}

\begin{figure*}[b!] 
    \centering
    \includegraphics[width=0.8\linewidth]{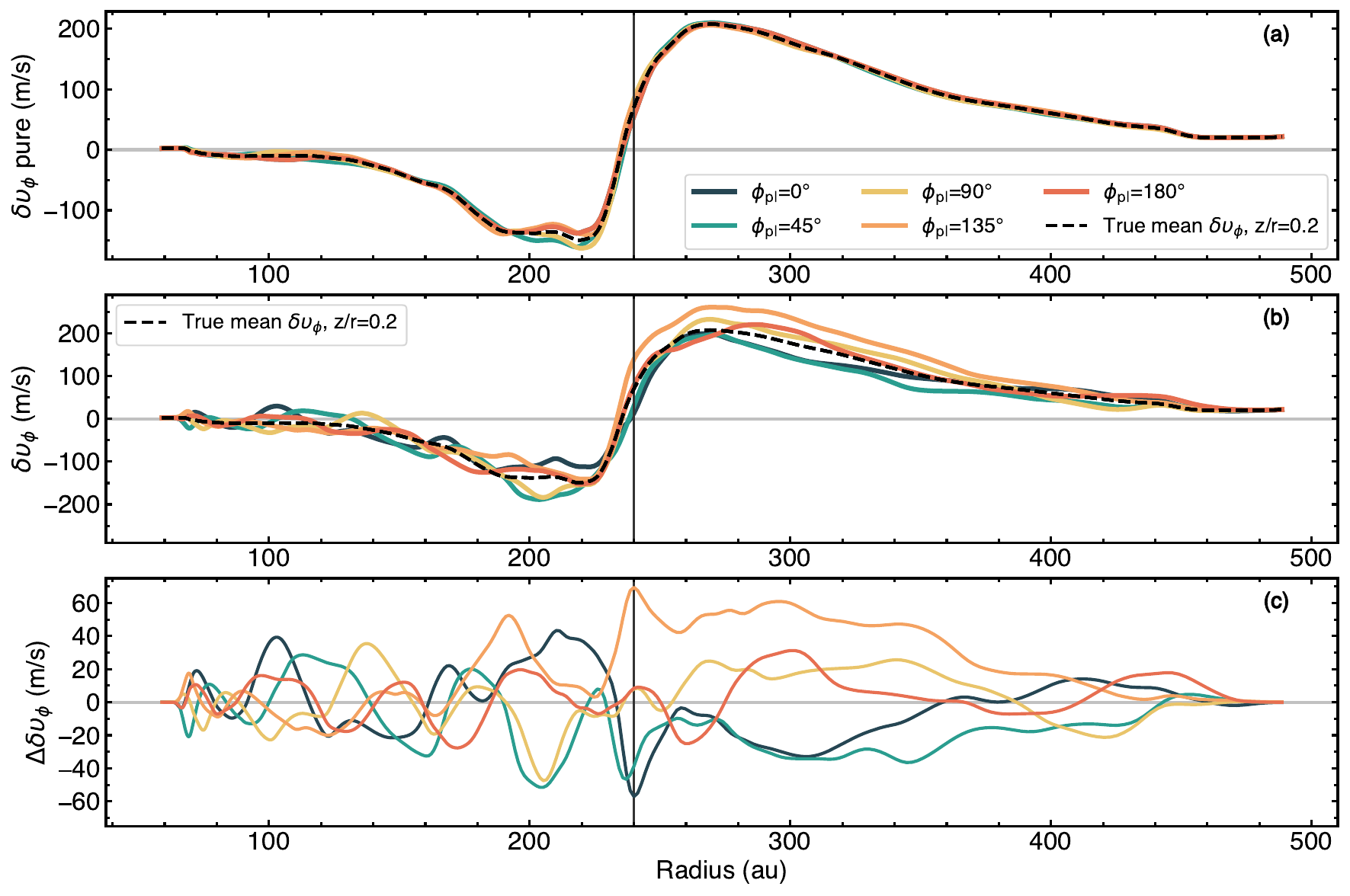}
    \caption{Radial profiles of deviations from Keplerian rotation extracted from a 3D hydrodynamical simulation \citep[][]{Teague_ea_2019a}. The dashed black true mean \dvphi{} line is taken directly from the hydrodynamical simulation. The colored \dvphi{} lines are extracted from the hydrodynamical data, projected onto the sky, and then applied the discminer approach (see Eq.\,\ref{eq:vphi_dm}. The different lines correspond to a planet at different azimuthal locations ($\phi_{\rm pl}$) in the disk measured counterclockwise from the red-shifted disk major axis.}
    \label{fig:planet_azimuth}
\end{figure*}

In this section of the appendix, we want to assess biases in the principal component decomposition of our line-of-sight velocity components. In particular, in the case of a massive embedded planet, how azimuthal velocities can be mislead for radial velocities and vice versa. To this means we use a 3D hydrodynamical model with the identical setup to the one presented in \cite{Teague_ea_2019a} with one embedded two Jupiter mass planet located at 240\,au.

We now want to assess how the excited density waves of the planet can introduce velocity perturbations in \vphi{} and \vrad{}, that break our assumption of the rotational velocity being azimuthally symmetric around the disk's minor axis and dominant over radial and vertical ones in the line-of-sight (see Eq.~\ref{eq:vphi_los}). First, we extract the three velocity components \vphi{}, \vrad{} and \vz{} from the simulation at $z/r=0.2$ and then project them onto the sky assuming a moderate disk inclination of 30\degree along our line-of-sight (see Eq.~\ref{eq:v_los}). Secondly, we apply Equation~\ref{eq:vphi_dm} to extract \vphi{} from this line-of-sight velocity. We can then compare our extracted \vphi{} to our true mean input $\upsilon_{\phi, \rm{true}}$ from the simulation. For clarity, we additionally subtract the underlying Keplerian rotation from both simulated and measured rotational velocities to show the deviations from Keplerian rotation. In the last step, we change the position of the planet on the sky by varying its azimuthal location ($\phi_{\rm{pl}}$) in the disk in steps of 45\degree starting counterclockwise from the red-shifted major axis of the disk. So for example at 90\degree it is located exactly at the disk minor axis, at 180\degree at the blue-shifted major axis, and for $\phi_{\rm{pl}}$=45\degree, 135\degree right in between both. 

First, we want to understand if the planet itself introduces asymmetric perturbations in \vphi{}. Therefore, we neglect the other two velocity components in the line-of-sight, which means setting \vrad{}=0=\vz{} (see Eq.\,\ref{eq:v_los}). In panel a) of Figure~\ref{fig:planet_azimuth}, we now show the extracted \vphi{} curves compared to the true mean of the simulation. We see that irrespectively of the planet's azimuthal location we retrieve the \dvphi{} almost perfectly. This means the \dvphi{} perturbations induced by the planet are inherently symmetric and do not introduce inaccurate \vphi{} velocity measurements.

Now, we consider all velocity components along the line-of-sight. Panel (b) shows the variations in the extraction of \dvphi{} introduced by wrongly attributing \vrad{} (less \vz{}) as \vphi{} and vice versa. This shows the shortcoming of our velocity extraction method, purely from projection effects along the line-of-sight. Panel (c) is the subtraction of the extracted \dvphi{}-curves from the true mean in panel (b). The strongest mismatch occurs for the planet being located right in between disk minor and major axis ($\phi_{\rm pl}=45\degree$ \& $\phi_{\rm pl}=135\degree$). These are the azimuths where \vphi{} and \vrad{} equally contribute to the line-of-sight velocity (compare Eq.\,\ref{eq:vphi_los}). At these locations, it is increasingly difficult to disentangle the velocity components and measure an accurate \vphi{}. Yet, these are also the azimuthal locations where it's easiest to detect planetary signatures in 2D residual maps, as has been extensively studied in \cite{Izquierdo_ea_2021}.

We conclude that even though the amplitude of \dvphi{} changes considerably for the planet at different azimuths, the overall shape of the \dvphi{} profile in (b) is nearly unaffected by this azimuthal dependency. Therefore, we conclude that retrieving the radial locations of pressure gaps through a radially increasing \dvphi{} profile is still robustly retrieved. We further note that radiative transfer effects, which we do not consider here for simplicity, can further complicate an accurate measurement of \vphi{}.

\newpage
\section{Complementary figures} \label{app:figures}

\begin{figure*}[h!]
    \centering
    \includegraphics[width=0.95\linewidth]{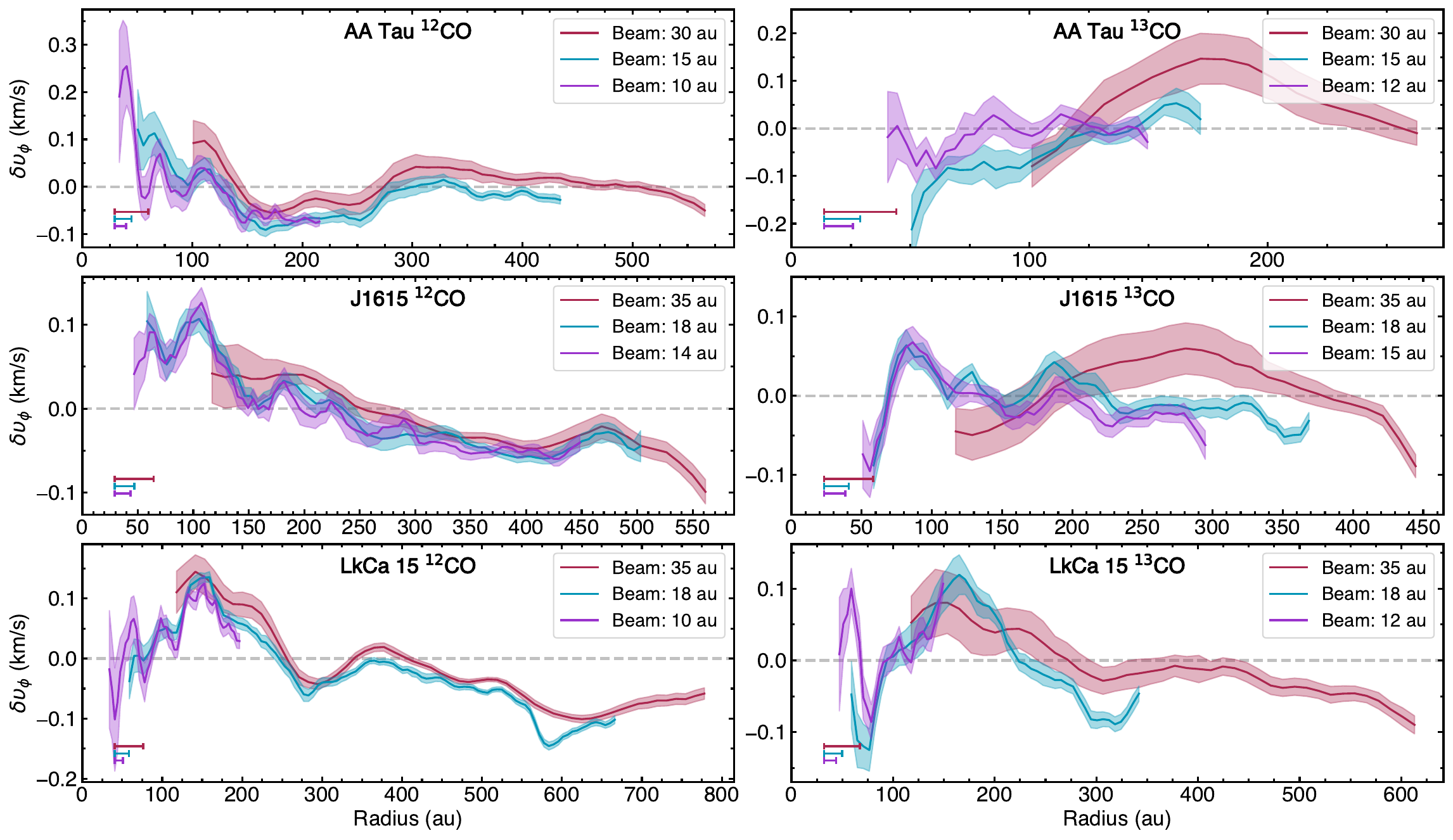}
    \caption{Smoothed radial profiles for \dvphi{} (\vphi{}-\vkep{}) for three different sources using the three sets of continuum-subtracted image cubes at various angular and spectral resolutions These cubes are namely the High Resolution Images (red, beamsize$<0.15\arcsec{}$), the Fiducial Images (turquoise, beamsize=0.15\arcsec{}), and High Surface Brightness Sensitivity Images (red, beamsize=0.30\arcsec{} \citep[see][for details]{Teague_exoALMA}. The colored shaded area of the lines shows the standard deviation within each
    extracted radial annulus. The beam sizes for each image cube are plotted in their respective color in the lower left corner. \\
    The plot shows that the choice of imaging parameters does not negatively affect the analysis products over the scales of interest. In our cases, these are either the innermost disk regions co-located with the continuum or the global scales tracing the outermost disk radii. The \dvphi{} profiles agree within their uncertainties. However, caution is necessary to ensure sufficient angular resolution for rotation curve analysis; otherwise, a beam that is not sufficiently small may smear out the substructures of interest, as can be seen in the overlapping profiles.}
    \label{fig:dvphi_var_resolutions}
\end{figure*} 

\begin{figure*}[h!]
    \centering
    \includegraphics[width=0.95\linewidth]{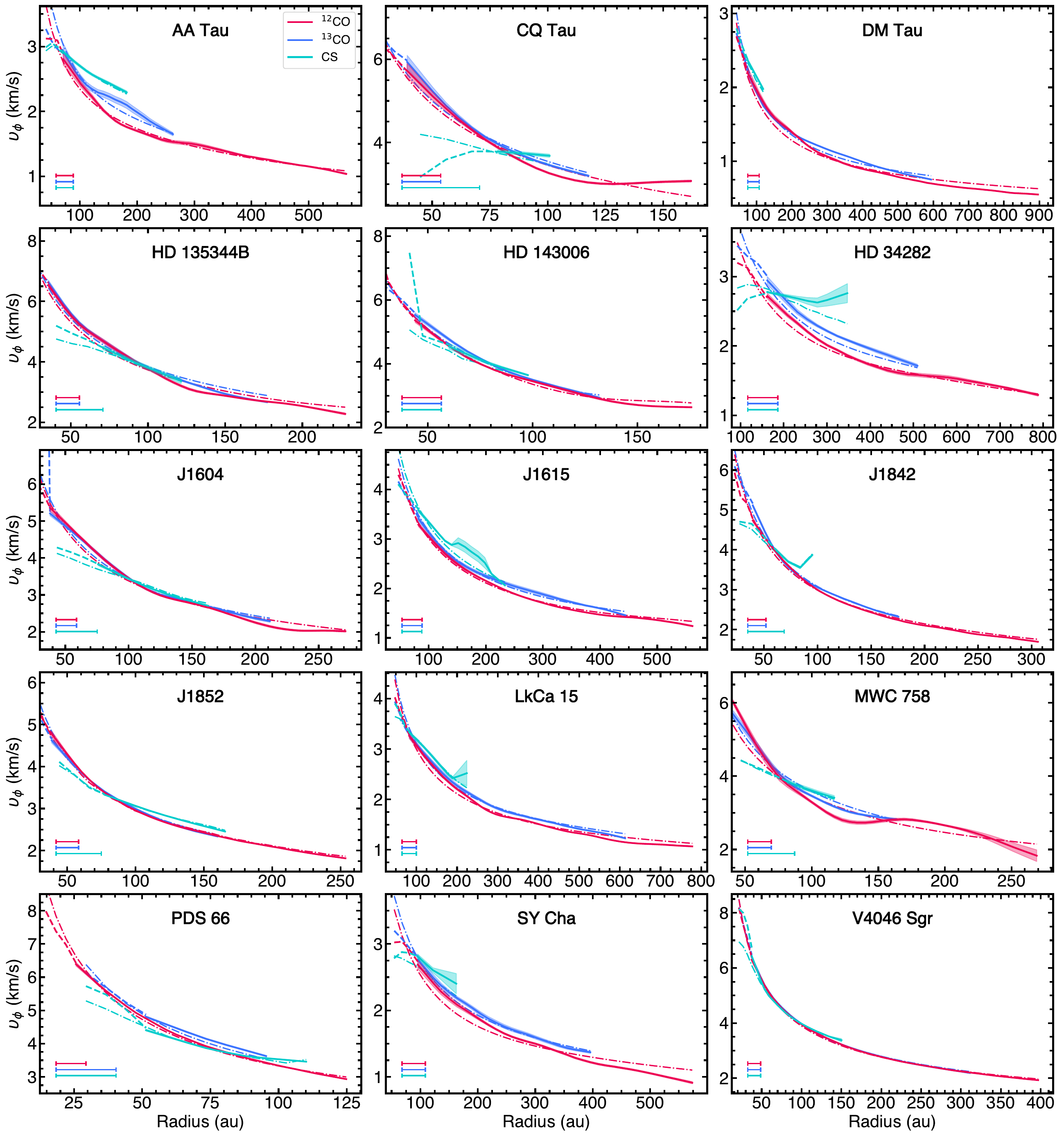}
    \caption{Rotation curves for all sources including the ones for \CSfull{}. For most sources, the CS rotation curves rotate faster than their CO counterparts, further providing evidence for vertical stratification since the CS molecular line emitting height is located closer to the colder midplane and thus shows faster rotation than the CO molecules located at the upper warm irradiated surface layers. The innermost two beam sizes from the center are plotted with dashed lines due to uncertainties in the velocity extraction. The colored shaded area of the lines shows the standard deviation within each extracted radial annulus. The reference Keplerian rotation is plotted as a thin dashed-dotted line for each molecule. The beam sizes for each molecular line are plotted in their respective color in the lower left corner.}
    \label{fig:rot_curves_CS}
\end{figure*} 

\begin{figure*}[t]
    \centering
    \includegraphics[width=1\linewidth]{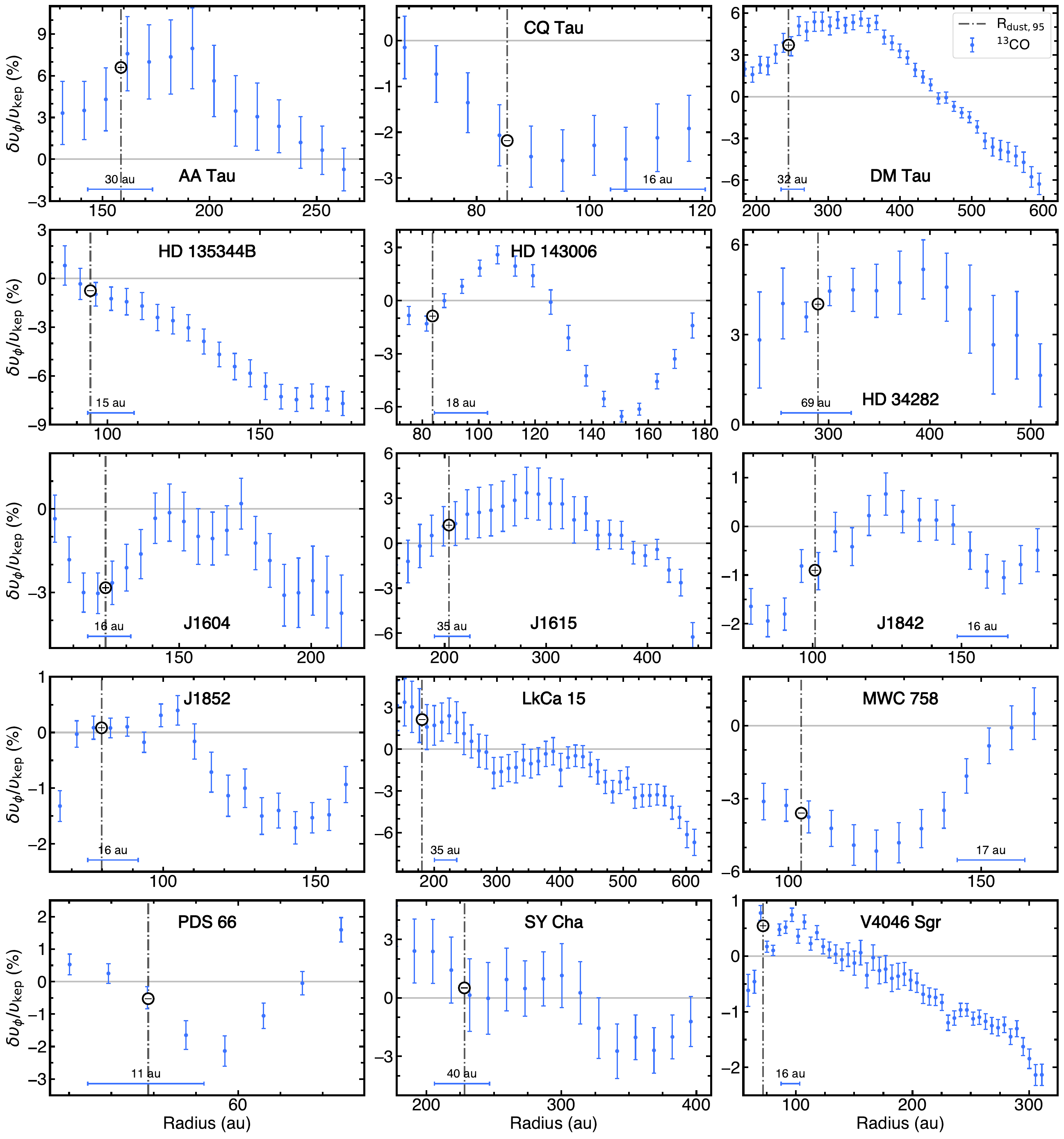}
    \caption{Radial profiles of \dvphi{} for \thCOfull{}{} of all sources focused on the region beyond the continuum substructures using the High Surface Brightness Sensitivity Images. The error bars show the standard deviation within each extracted radial annulus. The vertical dashed-dotted line indicates the radius that encompasses 95\% of the continuum emission, with the signs of the radial derivatives of \dvphi{} marked at this location. The beam size is shown in the lower left corner.}
    \label{fig:dvphi_large_13co}
\end{figure*}

\begin{figure*}[t]
    \centering
    \includegraphics[width=1\linewidth]{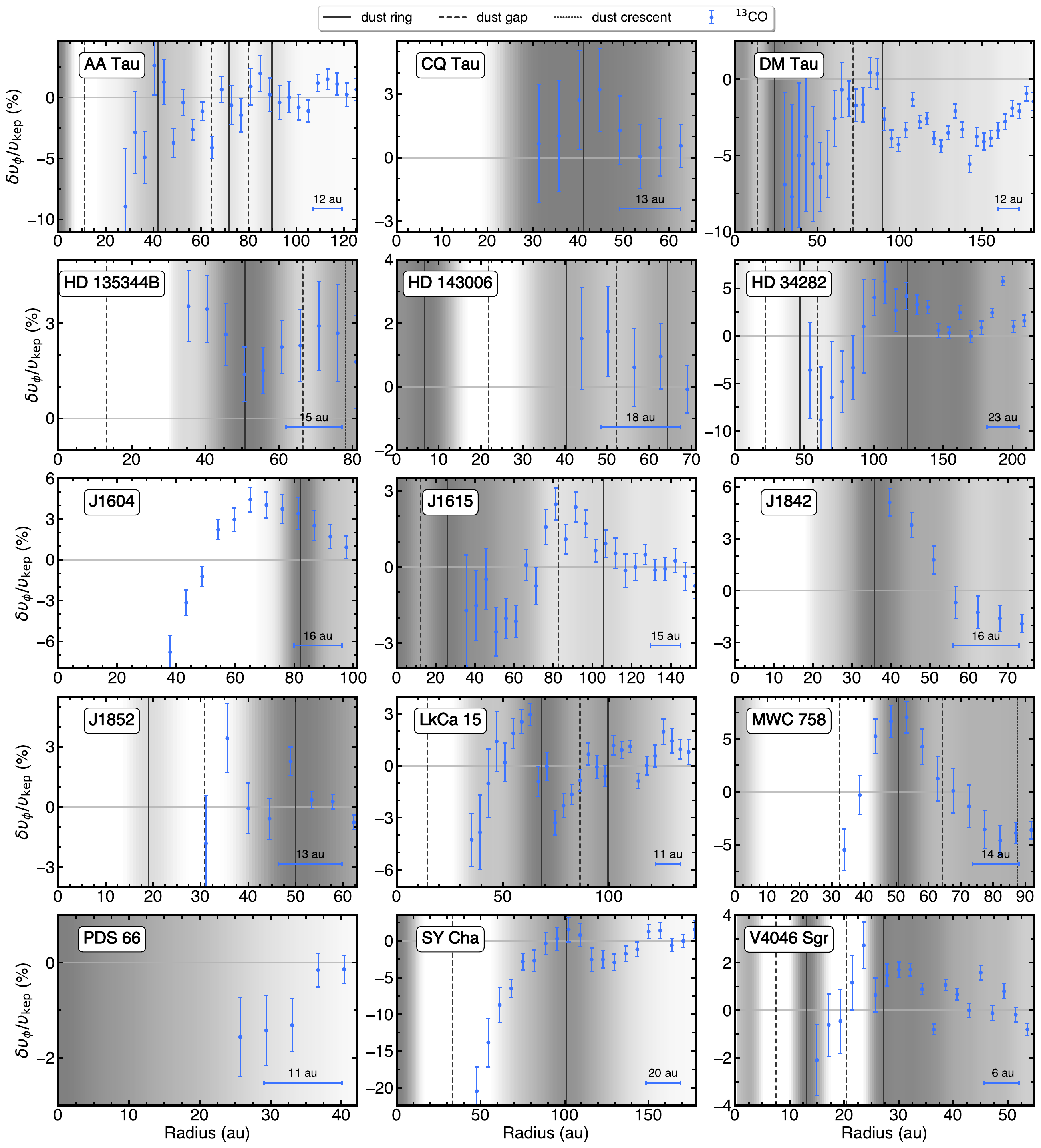}
    \caption{Radial profiles of \dvphi{} for \thCOfull{} of all sources focused on the region of the continuum substructures, using the High Resolution Images. The profiles are plotted starting at two beam sizes from the disk center and the error bars show the standard deviation of each bin. The gray background gradient highlights the \texttt{frank} radial profiles of the dust continuum emission normalized to its peak. The locations of dust rings, gaps, and crescents are plotted in solid, dashed, and dotted vertical gray lines, respectively \citep[][]{Curone_exoALMA}. If caused by pressure maxima and minima then we expect dust rings and gaps to be co-located with a decreasing and increasing \dvphi{} profile respectively. The beam size is plotted in the lower right corner.}
    \label{fig:dvphi_small_13co}
\end{figure*}

\begin{figure*}[t]
    \centering
    \includegraphics[width=1\linewidth]{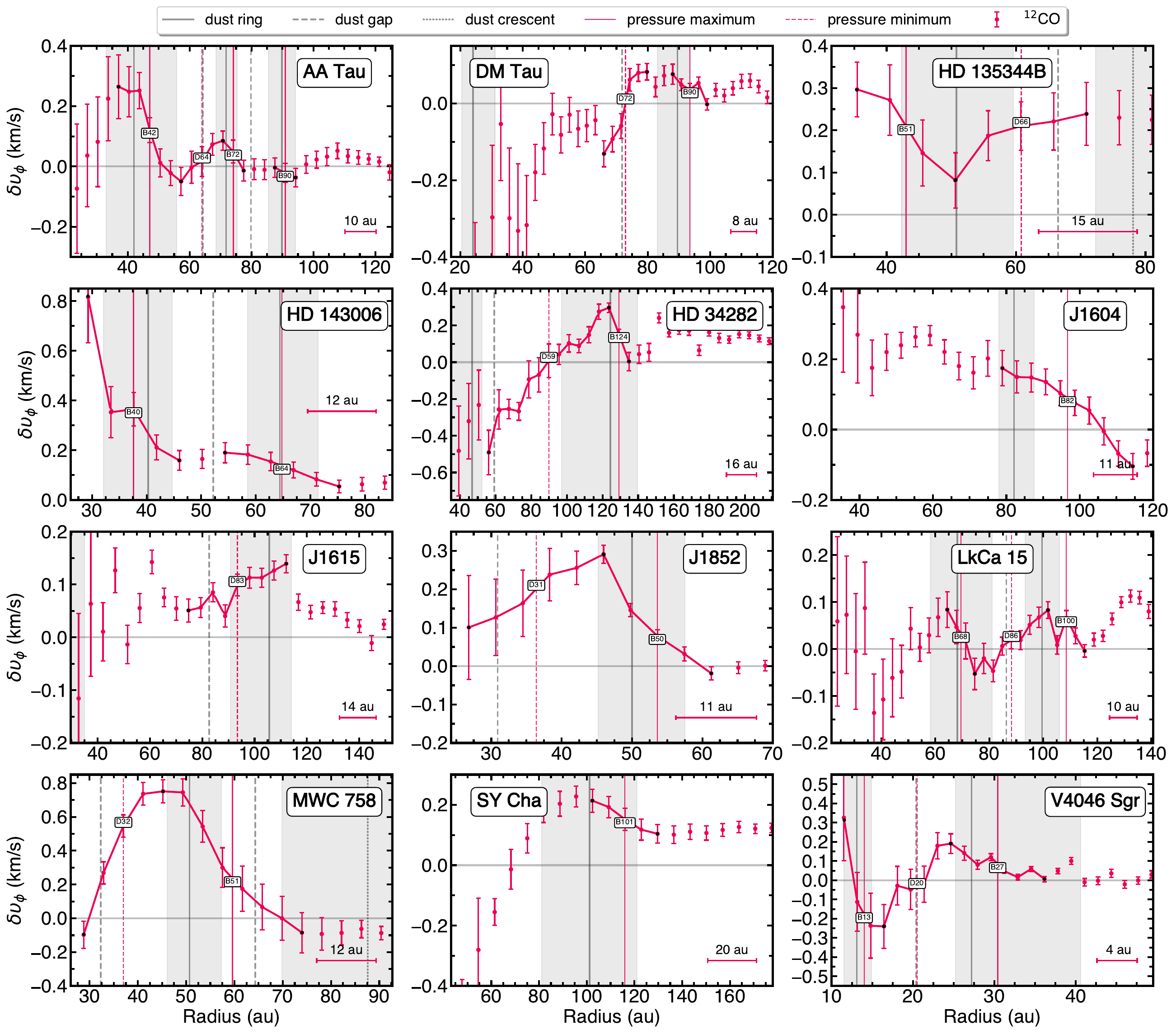}
    \caption{\twCO{} radial profiles of \dvphi{} for sources with pressure substructures co-located with the continuum substructures using the High Resolution Images. \dvphi{} substructures are labeled in each subplot as listed in Table~\ref{tab:substructures}. The start and end points of the substructures are annotated with black dots with connecting lines in between them. The vertical dashed and solid red lines indicate the centers' of pressure minima and maxima, respectively. The vertical grey shaded areas show the width of the continuum rings \citep[][]{Curone_exoALMA}. Additional plot annotations are the same as in Figure~\ref{fig:dvphi_small_13co}.}
    \label{fig:dvphi_substr_12co}
\end{figure*}

\begin{figure*}[t]
    \centering
    \includegraphics[width=1\linewidth]{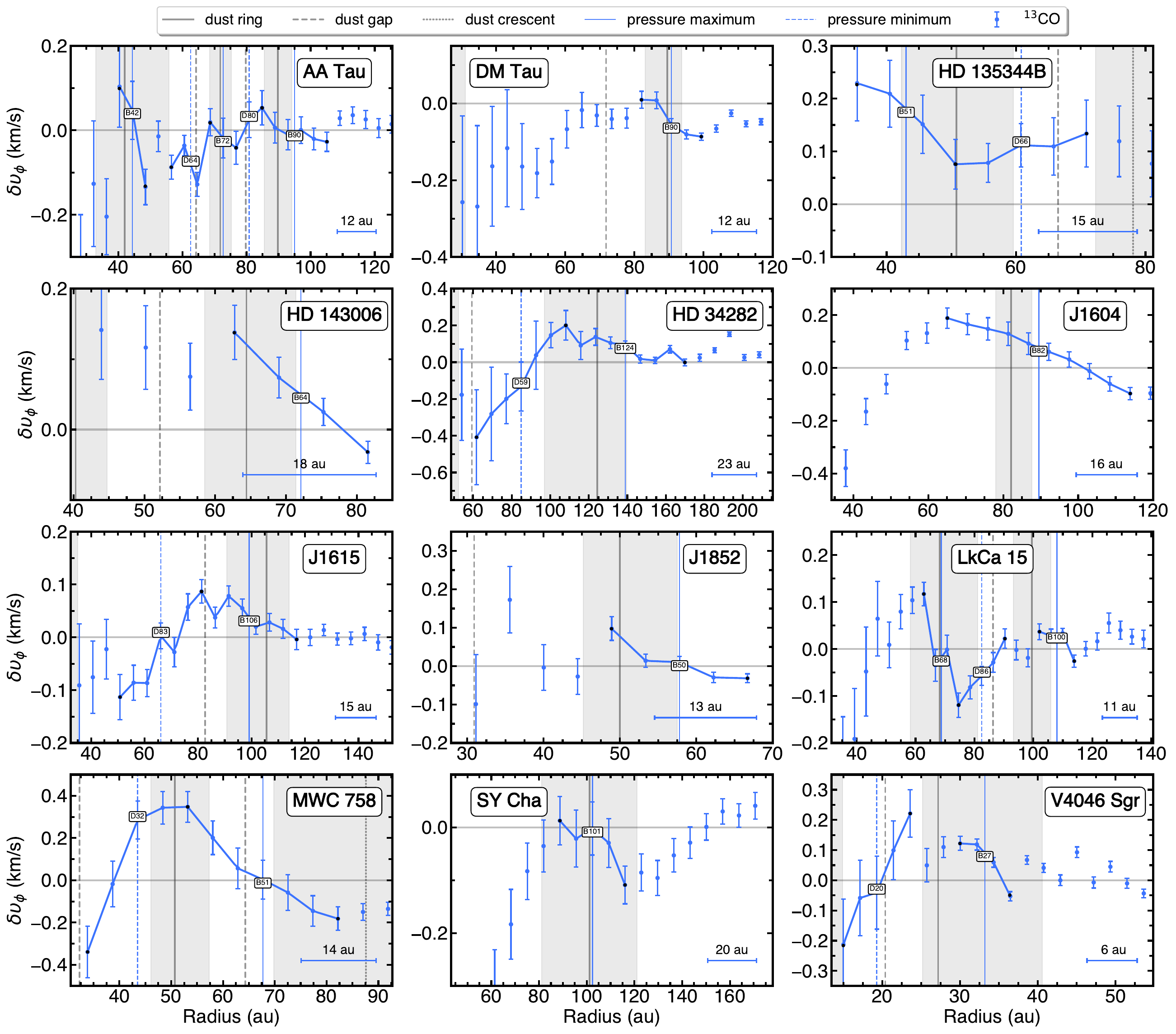}
    \caption{\thCO{} radial profiles of \dvphi{} for sources with pressure substructures co-located with the continuum substructures using the High Resolution Images. \dvphi{} substructures are labeled in each subplot as listed in Table~\ref{tab:substructures}. The start and end points of the substructures are annotated with black points with connecting lines in between them. The vertical dashed and solid blue lines indicate the centers' of pressure minima and maxima, respectively. The vertical grey shaded areas show the width of the continuum rings \citep[][]{Curone_exoALMA}. Additional plot annotations are the same as in Figure~\ref{fig:dvphi_small_13co}.}
    \label{fig:dvphi_substr_13co}
\end{figure*}
\begin{figure*}[h!]
    \centering
    \includegraphics[width=1\linewidth]{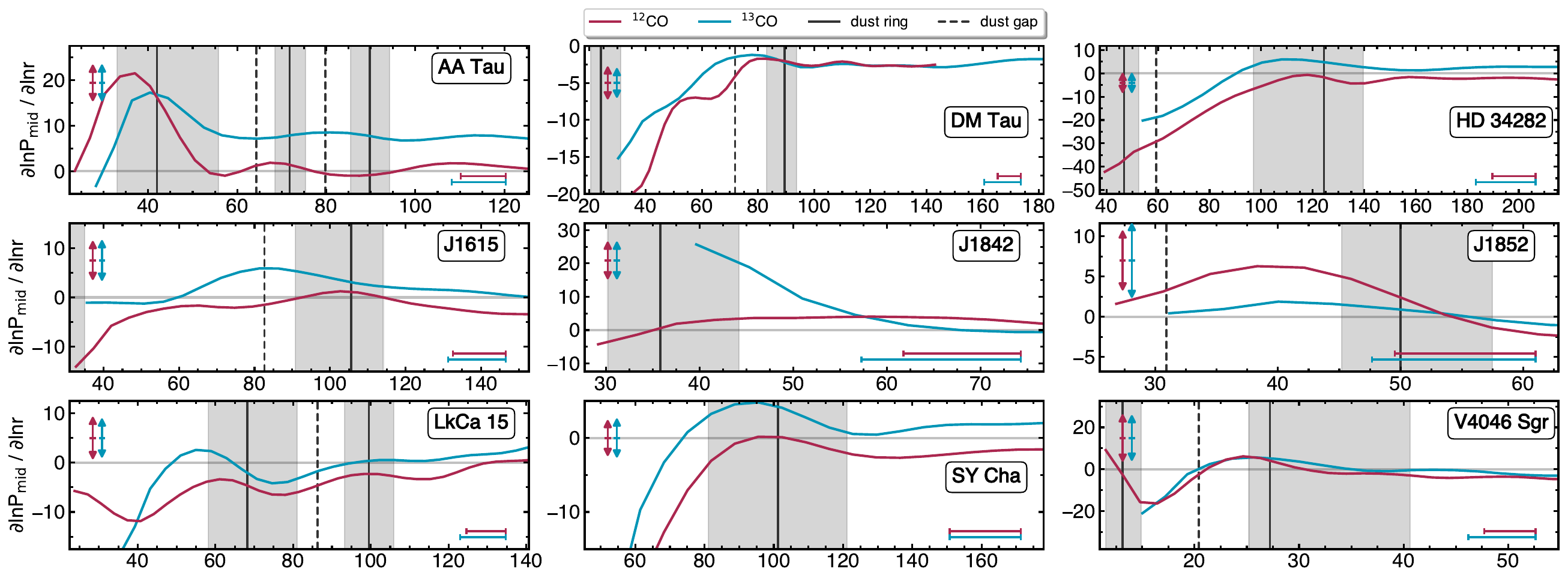}
    \caption{Radial profiles of the logarithmic midplane pressure derivative for all sources with temperature profiles focused on the region of the continuum emission using the High Resolution Images. The vertical colored arrows in the upper left show variations in the stellar mass of $\pm$3\,\% which would result in a shift of the whole profiles up and down. We expect dust rings and gaps to be co-located with \dpmid{}=0 if induced by pressure maxima and minima (see Fig.~\ref{fig:vphi_press}~b). The vertical grey shaded areas show the width of the continuum rings \citep[][]{Curone_exoALMA}. The beam sizes for each molecule are plotted in the lower right corner in their respective color.}
    \label{fig:pressure_small}
\end{figure*}

\begin{figure*}[h!]
    \centering
    \includegraphics[width=1\linewidth]{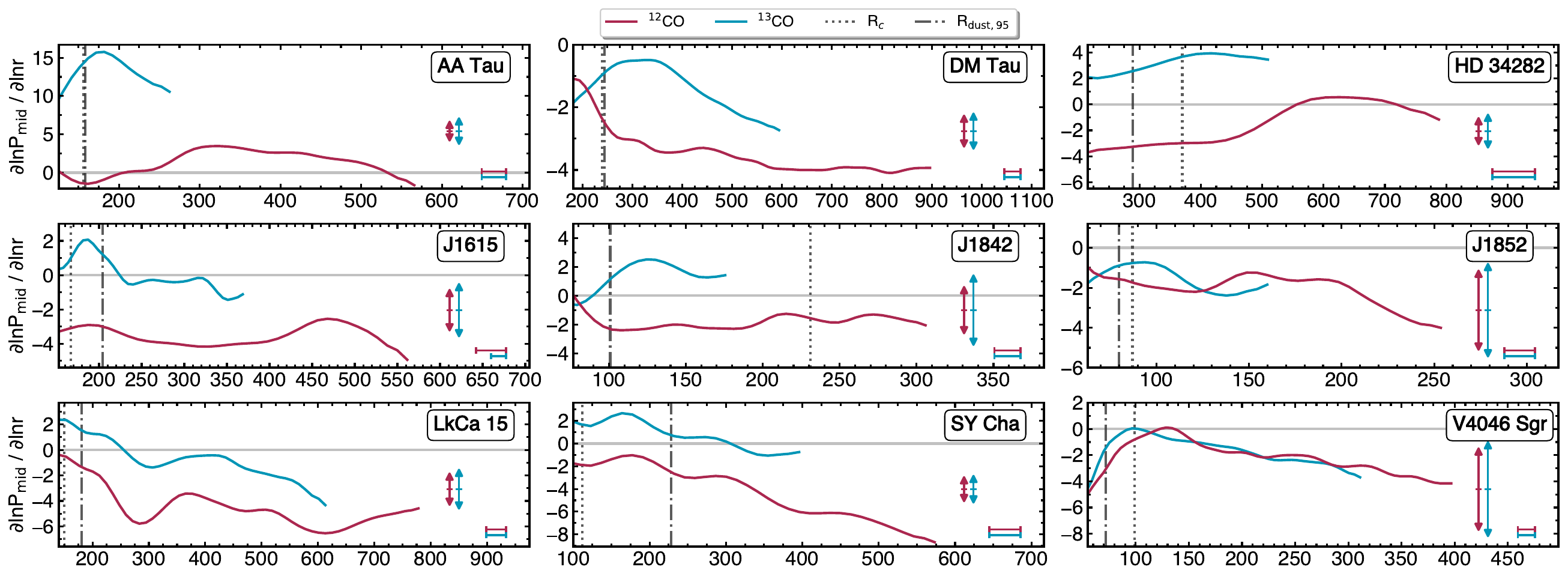}
    \caption{Same as Figure~\ref{fig:pressure_small} but focused on the disk region beyond the continuum emission using the High Surface Brightness Sensitivity Images. The dashed-dotted vertical line shows the radius enclosing 95\% of the continuum emission \citep{Curone_exoALMA} and the dotted vertical line indicates the scale radius $R_c$ for a \cite{Lynden_Bell_Pringle_1974} $\Sigma$-profile derived by \cite{Longarini_exoALMA}}
    \label{fig:pressure_large}
\end{figure*}

\begin{figure*}[h!]
    \centering
    \includegraphics[width=1\linewidth]{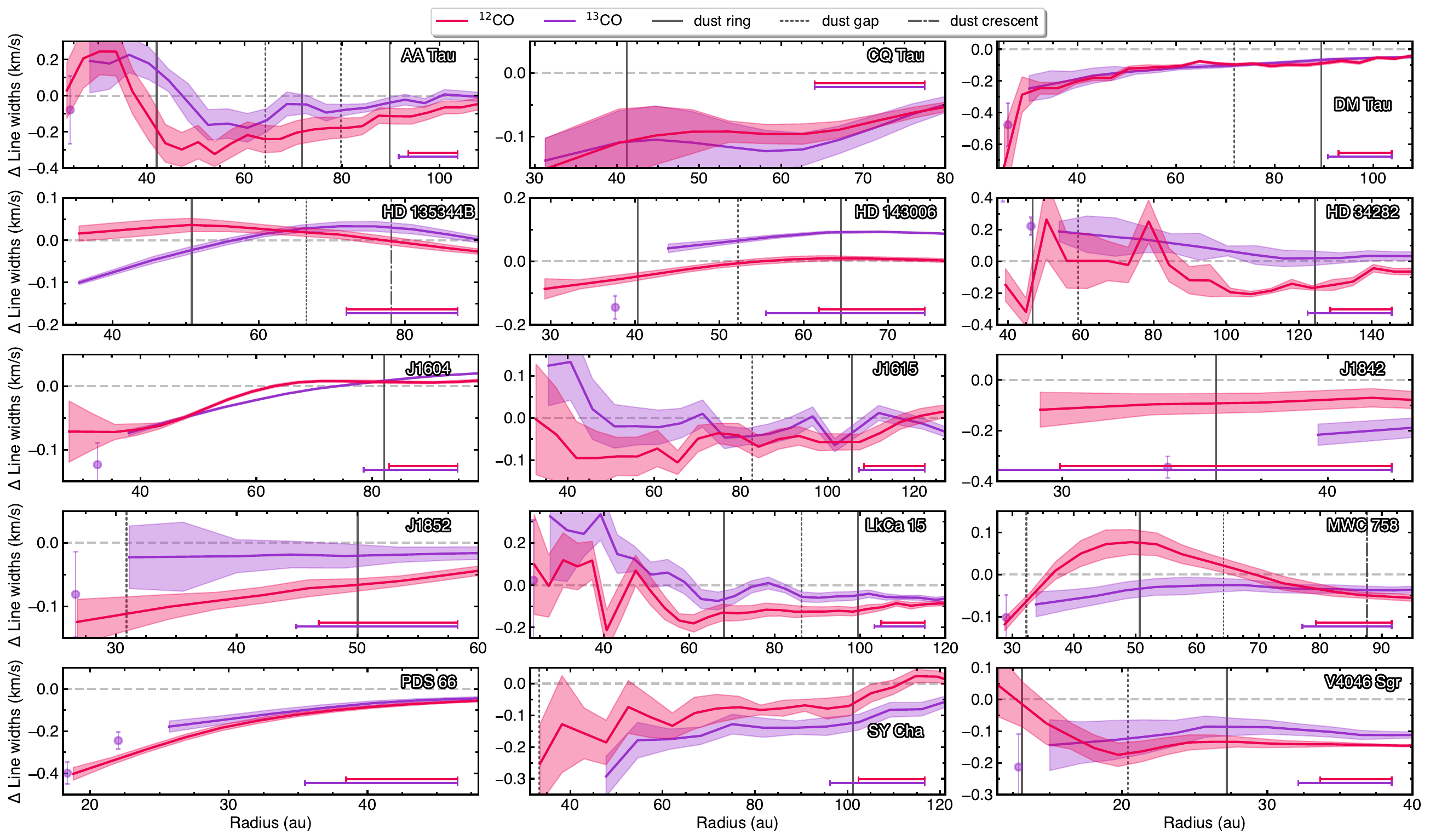}
    \caption{Radial profiles of the CO line width residuals for the whole sample focused on the region of the continuum emission using the High Resolution Images.}
    \label{fig:delta_linewidth}
\end{figure*}

\begin{figure*}[h!]
    \centering
    \includegraphics[width=1\linewidth]{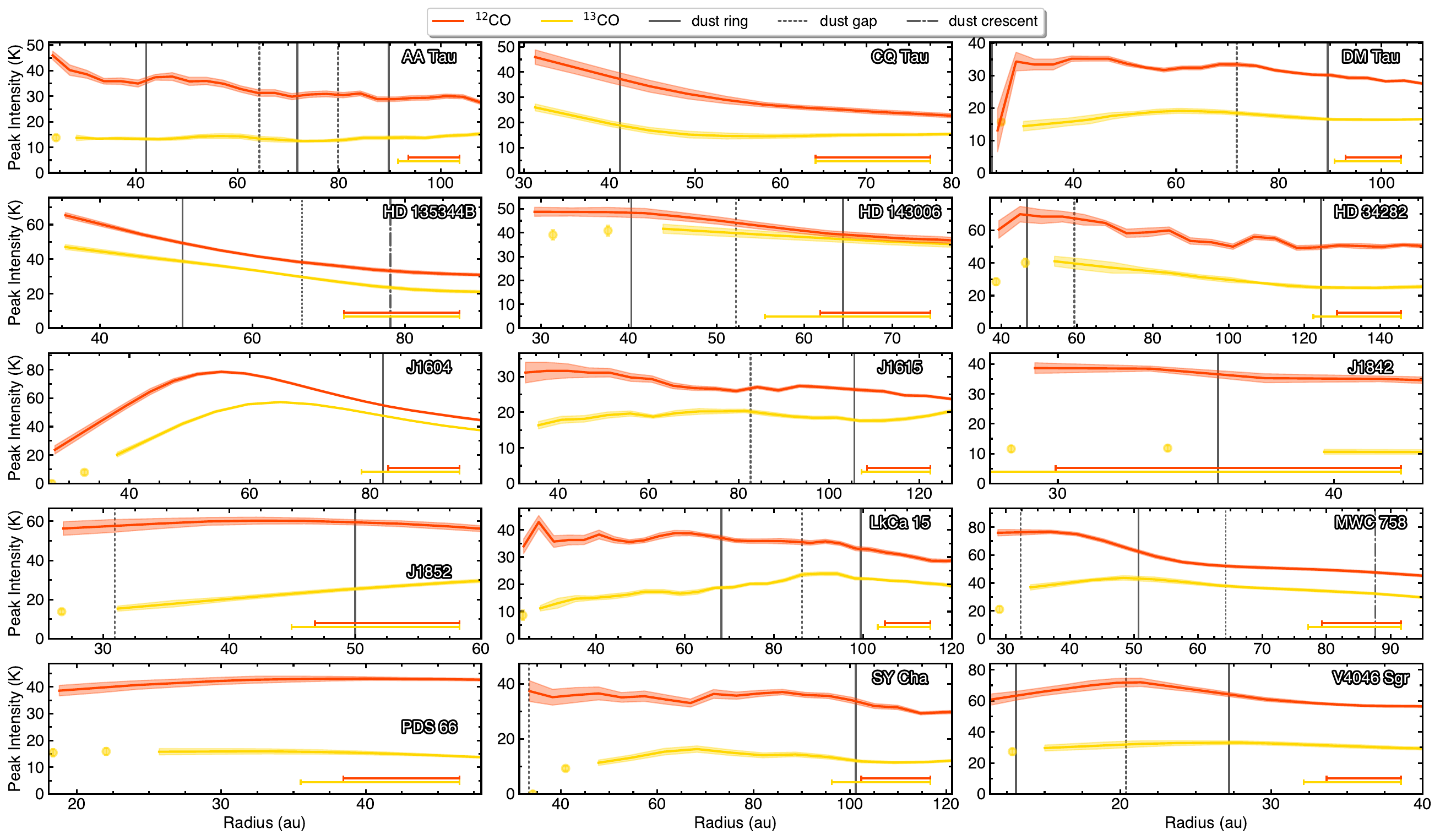}
    \caption{Radial profiles of the CO peak intensities for the whole sample focused on the region of the continuum emission using the High Resolution Images.}
    \label{fig:peakint}
\end{figure*}

\begin{figure*}[t]
    \centering
    \includegraphics[width=1\linewidth]{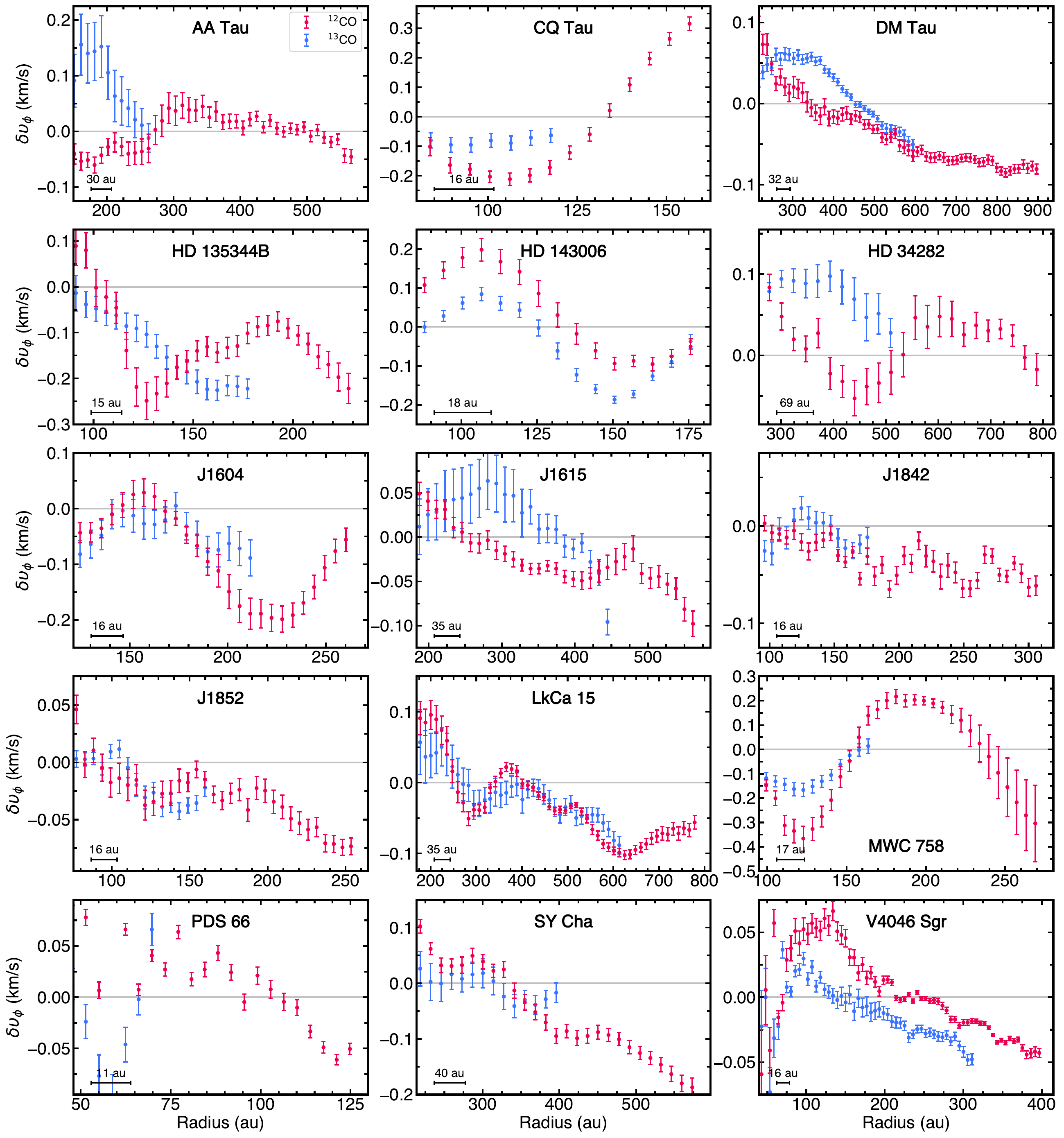}
    \caption{\twCO{} and \thCO{} radial profiles of \dvphi{} using the High Surface Brightness Sensitivity Images. Same as Figures ~\ref{fig:dvphi_large}~\&~\ref{fig:dvphi_large_13co}, but here showing \dvphi{} in units of km/s.}
    \label{fig:dvphi_large_kms}
\end{figure*}

\end{document}